\documentclass{pasj00}
\draft

\begin{document}
\SetRunningHead{Kanno, Harada \& Hanawa}{M1 Model}
\Received{2010/10/25}
\Accepted{2013/2/20}%{yyyy/mm/dd}
\Published{2013/8/25}%{yyyy/mm/dd}

\title{Kinetic Scheme for Solving M1 Model of Radiative Transfer}

%%% begin:list of authors
% Do NOT capitalize all letters in "textsc".
\author{Yuji \textsc{Kanno} and Tetsuya \textsc{Harada}} %
 % \thanks{Example: Present Address is xxxxxxxxxx}}
\affil{Department of Physics, Graduate School of Science, Chiba University,
1-33 Yayoi-cho, Inage-ku, Chiba 263-8522}
\email{kannoyj@astro.s.chiba-u.ac.jp}
\email{haradatt@astro.s.chiba-u.ac.jp}
\and
\author{Tomoyuki  {\sc Hanawa}}
\affil{Center for Frontier Science, Chiba University,
1-33 Yayoi-cho, Inage-ku, Chiba 263-8522}
\email{hanawa@cfs.chiba-u.ac.jp}
%%% end:list of authors

%%% Please use the following style in case that sorting by 
%%% affiliation is impossible. 
%
% \author{%
%   D-Firstname \textsc{D-Familyname}\altaffilmark{1}
%   E-Firstname \textsc{E-Familyname}\altaffilmark{1,2}
%   and
%   F-Firstname \textsc{F-Familyname}\altaffilmark{2}}
% \altaffiltext{1}{Address of Institute}
% \email{ddddd@xxx.xxx.xx.xx}
% \email{eeeee@xxx.xxx.xx.xx}
% \altaffiltext{2}{Address of Institute}

%% `\KeyWords{}' always has to be placed before `\maketitle'.
\KeyWords{radiative transfer, scattering, stars: pre-main sequence} %Do NOT move this preamble from here!

\maketitle

\begin{abstract}
We show a numerical scheme to solve the moment equations of the
radiative transfer, i.e., M1 model which follows the evolution of 
the energy density, $ E $, and the energy flux, $ \mbox{\boldmath$F$} $.
In our scheme we reconstruct the intensity from $ E $ and $ \mbox{\boldmath$F$} $
so that it is consistent with the closure relation,
relation, $ \chi~=~(3 \, + \, 4 f ^2)/(5 \, + \, 2 \sqrt{4 \, - \, 3 f ^2} ) $.
 Here the symbols, $ \chi $, 
$ f~=~|\mbox{\boldmath$F$}|/(cE) $, and $ c $,  
denote the Eddington factor, the reduced flux, and the speed of light, respectively.    
We evaluate the numerical flux across the cell surface from the kinetically 
reconstructed intensity.    It is an explicit function of
$ E $ and $ \mbox{\boldmath$F$} $ in the neighboring cells across the
surface considered.     
We include absorption and reemission within a numerical cell in the
evaluation of the numerical flux.  The numerical flux approaches to
the diffusion approximation when the numerical cell itself is optically thick.
Our numerical flux gives a stable solution even when some regions
computed are very optically thick.   
We show the advantages of the numerical flux with examples.
They include flash of beamed photons and irradiated protoplanetary disks. 
\end{abstract}

\section{Introduction}

Radiation plays important roles in many astronomical objects and
media.   Thus we need to include the effects of radiation somehow
in a realistic numerical simulation.   However it is still an extremely
heavy load to solve the full radiative transfer, i.e., taking account of
energy spectrum and angular distribution as well as spatial distribution
and temporal evolution.    This is simply because the radiative intensity
is a function in 6 dimensional phase space.  We need to reduce the 
load by using some approximations.

There exists several types of ideas for reducing the computation
cost.  Each of them have advantages and weak points.  

 First we can use a symmetry to reduce the cost for computation.
If the spherical symmetry is a good approximation, the intensity is a function
in three dimensional phase space and the radiative transfer is relatively
easy to solve.  However this idea can be applied only to highly symmetric
systems.  

Second we can assume that only a few sources are dominant in the
radiation fields.   If most of light comes from restricted relatively small
number of sources, solving radiative transfer is relatively easy. 
 We need to solve only the light rays from the sources.   The number 
of light rays to be solved are greatly reduced.
It is a good approximation when
some stars or black holes are dominant light sources (see, e.g.,
\cite{susa06,okamoto12}, and the references therein).    
However reflection and reemission from diffuse media cannot be taken
into account under this approximation.

Third we can use the moments of radiative intensity such as the energy
density by using some approximation on the angular distribution such
as the flux limited diffusion (FLD, see  e.g., \cite{levermore84}).   
 FLD uses only the radiative
energy density, which is equivalent to the average radiative intensity over
the solid angle of photon traveling, to express the radiation field.
The radiative flux is derived from the gradient of the energy density.
This approximation is good when the medium is optically thick and
radiation field is nearly isotropic.   It is not so bad even when the medium
is transparent.   However, this approximation cannot express a shadow,
a dark region behind an opaque object of finite extent.   Spurious 
radiation erases out the shadow in FLD \citep{gonzalez07}.  
Scattering is another weak  point of FLD.   Scattering changes angular
distribution of intensity but not the energy density.   Thus FLD cannot
evaluate the effects of scattering since FLD takes account of only
energy density.

\citet{gonzalez07} proposed M1 model in which the radiation field is
expressed by the energy density and flux, i.e., both the 0th and 1st
moments of the radiative intensity.  It is demonstrated that M1 scheme
can simulate a shadow by an opaque sphere successfully.  
M1 scheme can take account of scattering.  Scattering reduces the
energy flux while keeping the radiation density constant.
When the scattering is isotropic, the effect is correctly taken into account
in M1 model.  However, also M1 model has several weak points
and limitations.   M1 model cannot solve crossing of two beamed 
lights.   They erroneously merge into a beam at the crossing 
point.   This limitation comes from the fact that M1 model evaluates
higher moments of radiation intensity from the 0th and 1st moments.
In other words, M1 model has the lowest angular resolution and crossing
of two beams are beyond the scope.   This limitation is compensated 
by the low computation cost.   

M1 model equations are similar to the  hydrodynamical equations 
in the conservation form.   Both the equations are hyperbolic and 
have source terms.   Thus we can apply numerical methods for
integration of the hydrodynamical equations to solve M1 model equations.  
However we have two concerns when solving M1 model equations.   First 
the characteristic speeds are complex and not easy to evaluate, though
modern schemes for numerical hydrodynamics rely on them
(see, e.g., \cite{toro09}).     We can avoid the
computation of  characteristics by using HLLE flux  but
the resultant flux is diffusive and makes a shadow dim  
as pointed out by \citet{gonzalez07}.   

Second, absorption and emission (the source terms) are dominant when
optically thick.   On the other hand,  the source terms due to gravity are
minor contributions in numerical hydrodynamics.   Thus they are simply
added after solving wave propagation.  This approach does not work well
in M1 model equations when a cell itself is optically thick.  
This difficulty is known as  the diffusion limit behavior and several
solutions are proposed in the literature (see. e.g., \cite{audit02,berthon07}).

In this paper we propose an idea to construct a numerical flux for M1 model
which is less diffusive and yet stable in the diffusion limit.   
First we show a method to evaluate a numerical
flux of M1 model from radiation intensity kinetically reconstructed from 
the radiation energy density and flux.   The reconstructed radiation intensity
is consistent with the closure relation, i.e., the formula to close the
moment equations of the radiative transfer.  We evaluate  the radiative 
flux and pressure across the boundary between two adjacent 
computation cells by integrating the reconstructed intensity over the
solid angle.   We use the intensity of the upwind side, the numerical
flux is subject to causality.  Fortunately, the numerical flux is an
explicit function of the radiation energy density and flux.
A similar scheme is constructed for gas dynamics and called \lq \lq
kinetic scheme'' (see, e.g., \cite{pullin80,deschpande86} and the
references cieted in \cite{hauck11}).  Thus we use the same
terminology in this paper.

Second we include absorption within a computation cell.  
The numerical flux is evaluated on the cell surface,
while the energy density and flux are evaluated at the cell center.
They can be appreciably different if the cell itself is optically thick.   We 
propose an interpolation formula which provides a good approximation
both in the optically thin and thick limits.   It approaches to one obtained
from the reconstructed intensity in the optically thin limit, while it 
does to the diffusion approximation in the optically thick limit.
We show that this numerical flux gives a stable solution even when 
the computation box contains both optically thin and thick cells.

This paper is organized as follows.  We describe our numerical
methods to solve the M1 model in \S 2.  We perform some simple
examples to show the nature of our numerical scheme in \S 3.
In \S 4, we apply M1 model to irradiated protoplanetary disks.  
We discuss accuracy and stability of our numerical flux in \S 5.
We also discuss affinity of M1 model for massively parallel computing
in \S 5.    Methods for constructing numerical flux of the second order
accuracy in space are given in Appendix.

\section{M1 Model}

\subsection{Basic Equations}

First we review M1 model of \citet{gonzalez07}.    We assume that 
emission is thermal and scattering is isotropic.  Then the radiative transfer
equation for the specific intensity, 
$ I _\nu \left( \mbox{\boldmath$x$},~t;~\mbox{\boldmath$n$} \right) $,
is expressed as
\begin{eqnarray}
\left( \frac{1}{c} \, \frac{\partial}{\partial t} \; + \;
\mbox{\boldmath$n$} \cdot \mbox{\boldmath$\nabla$} \right)
 I  _\nu \left( \mbox{\boldmath$x$},~t;~\mbox{\boldmath$n$} \right) 
  & = &  \kappa_{\nu,a} B _\nu \rho \left( \mbox{\boldmath$x$},~t \right)  \; - \;
  \left(\kappa _{\nu,a} \, + \, \kappa _{\nu,s} \right) \, \rho \, 
  I _\nu \left( \mbox{\boldmath$x$},~t;~\mbox{\boldmath$n$} \right) \nonumber \\
& + &
\kappa _{\nu,s} \rho \int   I _\nu \left( \mbox{\boldmath$x$},~t;~\mbox{\boldmath$n$}^\prime
 \right) \, d\mbox{\boldmath$n$} ^\prime \; , \label{transfer0}
 \end{eqnarray}
 where $ \rho $ and $ c $ denote the density the speed of light, respectively; 
 $ \kappa _{\nu,a} $  and $ \kappa _{\nu,s} $ 
 denote the absorption and scattering opacities at the photon frequency,
 $ \nu $, respectively.   The symbols, $ \mbox{\boldmath$n$} $ and
 $ \mbox{\boldmath$n$} ^\prime $, denote the angular variable, i.e., the
 unit vector parallel to the light propagation.  The symbol, $ B _\nu $, denotes
 the Planck function and is a function of the temperature, $ T $.
 
  We integrate Equation (\ref{transfer0}) over the
 solid angle to obtain
 \begin{equation}
 \frac{\partial E _\nu}{\partial t} \; + \;
\sum _{i=1} ^3
 \frac{\partial F _{\nu,i}}{\partial x _i}  \; =  \; \kappa _{\nu,a} \rho  
 \left( 4 \pi B _\nu \, -  \,  c E _\nu \right) \, , \label{moment0}
\end{equation}
where
\begin{eqnarray}
E _\nu \left( \mbox{\boldmath$x$},~t \right) & = & \frac{1}{c} \, \int I _\nu 
\left(\mbox{\boldmath$x$},~t;~\mbox{\boldmath$n$} \right) \, 
d \mbox{\boldmath$n$} \, ,  \\
F _{\nu,i}  \left( \mbox{\boldmath$x$},~t \right) & = & \int \left( \mbox{\boldmath$e$}_i \cdot 
\mbox{\boldmath$n$} \right)  \, I _\nu (\mbox{\boldmath$n$}) \, d \mbox{\boldmath$n$}  \, . 
\end{eqnarray}
The symbol, $ i $, specifies a direction in the Cartesian coordinates 
and $ \mbox{\boldmath$e$} _i $ does
the unit vector in the direction.
Equation (\ref{moment0}) denotes the conservation of radiation energy density,
$ E _\nu $, when the right hand side vanishes.   For simplicity we assumed that
the emission and absorption are isotropic.

Similary we obtain
\begin{equation}
\frac{\partial F _{\nu,i}}{\partial t} \; + \; c ^2 \,
\sum _{j=1} ^3  \frac{\partial P _{\nu,ij}}{\partial x _j} \; = \;
- \, c \, \left( \kappa _{\nu,a} \, + \, \kappa _{\nu,s} \right) \, \rho \,  F _{\nu,i} \, ,
\label{moment1}
\end{equation}
where
\begin{eqnarray}
P _{\nu,ij}  \left( \mbox{\boldmath$x$},~t \right) 
& = & \frac{1}{c} \, \int \, \left( \mbox{\boldmath$e$}_i \cdot 
\mbox{\boldmath$n$} \right) \, \left( \mbox{\boldmath$e$}_j \cdot 
\mbox{\boldmath$n$} \right) I _\nu (\mbox{\boldmath$n$}) \,d \mbox{\boldmath$n$} \, ,
\end{eqnarray}
by integrating Equation~(\ref{transfer0}) multiplied by $ \mbox{\boldmath$n$} $ 
over the whole solid angle.   The symbol, $ j $, as well as $ i $ specifies one
in the Cartesian coordinates.

When deriving Equation (\ref{moment1}),
we assumed that the scattering is symmetric with respect to forward and backward.
When the scattering is anisotropic, Equation (\ref{moment1}) should be replaced by
\begin{equation}
\frac{\partial F _{\nu,i}}{\partial t} \; + \; c ^2 \,
\sum _{j=1} ^3  \frac{\partial P _{\nu,ij}}{\partial x _j} \; = \;
- \, c \, \left[ \kappa _{\nu,a} \, + \, \kappa _{\nu,s} 
\left(1 \, - \, \langle \cos~\theta \rangle \right) \,
\right] \, \rho \,  F _{\nu,i} \,  ,
\label{moment1a}
\end{equation}
where $ \langle \cos~\theta \rangle $ denotes the scattering asymmetry parameter.

In order to solve Equations (\ref{moment0}) and (\ref{moment1}), we invoke the
closure relation,
\begin{eqnarray}
P _{\nu,ij}  & = & \left[
 \frac{1 \, - \, \chi _\nu}{2} \; + \; \left( \frac{3 \chi _\nu \, - \, 1}{2} \right)
\frac{f _{\nu,i} f _{\nu,j}}{|\mbox{\boldmath$f$} _\nu |^2} \, 
\right] \, E _\nu , \label{closure}
\end{eqnarray}
where
\begin{eqnarray}
\mbox{\boldmath$f$} _\nu & \equiv &  
\left(
\begin{array}{c}
f _{\nu,1} \\ f _{\nu,2} \\ f _{\nu,3} 
\end{array}
\right) \; = \;
\frac{1}{c E _\nu}  \, 
\left(
\begin{array}{c}
F _{\nu,1} \\ F _{\nu,2} \\ F_{\nu,3} 
\end{array} 
\right)  \, ,  \\
\chi _\nu  & = & \frac{3 \, + \, 4 | \mbox{\boldmath$f$} _\nu |^2}
{5 \, + \, 2 \sqrt{4 \, - \, 3 |\mbox{\boldmath$f$} _\nu  |^2}} \, .
\end{eqnarray} 

The above radiative transfer equation is solved in coupled with the 
hydrodynamical equations.   The solution of the hydrodynamical
equations provides the density, velocity, and temperature.  Hence
the opacity and source functions are evaluated as the functions of
them.   In this paper we consider only the change in the temperature
and neglect the changes in the density and velocity.   
This approximation can be justified when we consider the protoplanetary
disk in thermal equilibrium.   The frequency dependent opacity depends 
little on the density and temperature in a certain regime 
(see, e.g., \cite{henning96}), although the Rosseland mean opacity
does depend.

\subsection{Hydrodynamics}

The gas is heated by absorption and cooled by emission.   The heating and
cooling are evaluated by
\begin{equation}
\rho T \, \frac{D s}{D t} \; = \;
\int _0 ^\infty \sigma _{\nu,a} \, \left[ c E _\nu \, - \, 4 \pi \, B _\nu (T) \right]
\, d\nu \, ,
\end{equation}
where $D/Dt$ and $ s $ denote the Lagrange derivative and  the specific entropy,
respectively.  Then the hydrodynamical
equations are written in the conservation form as
\begin{equation}
\frac{\partial \mbox{\boldmath$U$} _H }{\partial t}  \; + \; 
\sum _{i=1} ^3 \frac{\partial \mbox{\boldmath$F$} _{H,i}}{\partial x _i} 
\; = \; \mbox{\boldmath$S$} _H \, ,
\end{equation}
where
\begin{equation}
\mbox{\boldmath$U$} _H  \; = \;
\left( \begin{array}{c}
\rho \\ \rho v _1 \\ \rho v _2 \\ \rho v _3 \\ \rho E _H
\end{array} \right) \, , \hskip 0.5cm
\mbox{\boldmath$F$} _{H,i} \; = \;
\left( \begin{array}{c}
\rho v _i \\
\rho v _1 v _i \, + \, P \delta _{i,1} \\
\rho v _2 v _i \, + \, P \delta _{2,i} \\
\rho v _3 v _i \, + \, P \delta _{3,i} \\
\rho H _H v _i
\end{array}
\right) \, , 
\end{equation}
\begin{equation}
\mbox{\boldmath$S$} _H \; = \;
\left\{
\begin{array}{c}
0 \\ \rho g_1 \\ \rho g_2 \\ \rho g _3 \\ \rho \mbox{\boldmath$v$} \cdot
\mbox{\boldmath$g$} \, + \,  
\int _0 ^\infty \sigma _{\nu,a} \, \left[ c E _\nu \, - \, 4 \pi \, B _\nu (T) \right]
\, d\nu \end{array}
\right\} \, ,
\end{equation}
\begin{eqnarray}
E  _H & = & \frac{|\mbox{\boldmath$v$} |^2}{2} \; + \;
\frac{1}{\gamma \, - \, 1} \, \frac{P}{\rho} \, , \\
H  _H & = & \frac{|\mbox{\boldmath$v$} |^2}{2} \; + \;
\frac{\gamma}{\gamma \, - \, 1} \, \frac{P}{\rho} \, . 
\end{eqnarray}
The symbol, $ \mbox{\boldmath$v$} \, = \, \left( v _1,~v_2,~v _3 \right) $, denotes
the gas velocity and the symbol, $ \mbox{\boldmath$g$} $ = 
$ (g _1,~g_2,~g _3) $, does the gravity.
The gas is assumed to be ideal one of which the specific heat ratio is $ \gamma $.

\subsection{Numerical Scheme}

Equations (\ref{moment0}) and (\ref{moment1}) can be expressed as
\begin{equation}
\frac{\partial \mbox{\boldmath$U$}_\nu}{\partial t} \; + \;
\sum _{i=1} ^3 \frac{\partial \mbox{\boldmath$F$}_{\nu,i}}{\partial x _i} \; 
= \; \mbox{\boldmath$S$} _\nu \; . 
\end{equation}
\begin{equation}
\mbox{\boldmath$U$} _\nu \;  = \; 
\left( \begin{array}{c}
E _\nu \\ F _{\nu,1} \\ F _{\nu,2} \\ F _{\nu,3} 
\end{array}
\right) \, , \hskip 0.5cm
\mbox{\boldmath$F$} _{\nu,i}  \; = \; 
\left( \begin{array}{c}
F _{\nu,i} \\ c ^2 P _{\nu,1i} \\ c ^2 P _{\nu,2i} \\ c ^2 P _{\nu,3i} 
\end{array}
\right) \, , \hskip 0.5cm
\mbox{\boldmath$S$}  _\nu \; = \;
\left[
\begin{array}{c}
\sigma _{\nu,a} \left( 4 \pi B _\nu \, - \, c \, E _\nu \right) \\
- \, c \, \left( \sigma _{\nu,a} \, + \, \sigma _{\nu,s} \right) \, F  _{\nu,1} \\
- \, c \, \left( \sigma _{\nu,a} \, + \, \sigma _{\nu,s} \right) \, F  _{\nu,2} \\
- \, c \, \left( \sigma _{\nu,a} \, + \, \sigma _{\nu,s} \right) \, F  _{\nu,3} 
\end{array}
\right] \, .  \label{conservation-form}
\end{equation}
Equation (\ref{conservation-form}) has the same structure as that of 
the hydrodynamical equations in the conservation form.
Thus the Godunov-type method, which is often used for solving the hydrodynamical
equations (see, e.g., \cite{toro09}), can be applied to Equation (\ref{conservation-form}).

In the Godunov-type method, the time evolution is evaluated based on the characteristics,
i.e., the propagation speeds of signal.    The characteristics of 
Equation (\ref{conservation-form}) is rather complex and the computation of them takes
much time.    \citet{gonzalez07} obtained them by interpolating the table prepared
rather than by computing them at each time step.   

We can avoid detailed computation of the characteristics by employing HLLE scheme,
in which we need only the upper  and lower limits on the characteristics 
(cf. \cite{toro09}).   However,  HLLE scheme gives us a too much diffusive solution
if the upper and lower limits are taken to be the speed of light, $ \pm c $
\citep{gonzalez07}.   

Ideal characteristics should be evaluated from the radiation fields in the
adjacent cells across the surface on which the numerical flux is evaluated.
Figure~1 of \citet{gonzalez07} shows the characteristics for a given 
radiation field.   The ideal characteristics should be appropriately
averaged ones when the radiation fields differ appreciably in the 
two adjacent cells.   \citet{roe81} obtained such average characteristics
for the hydrodynamical equations.   Similar average characteristics 
have not been obtained for the moment equation of the radiative transfer.

\subsection{Kinetic Reconstruction of the Intensity}

We obtain the numerical flux not by computing the characteristics but by
reconstructing the intensity consistent with the moments of the radiation
field.   When the intensity is expressed as
\begin{equation}
I_\nu(\mbox{\boldmath$n$}) \; = \;
\frac{3~c~(1~-\beta _\nu ^2)^3~E _\nu}{8 \pi \, (3~+~\beta _\nu ^2)}
\, \left(1~-~\mbox{\boldmath$\beta$} _\nu \cdot \mbox{\boldmath$n$} \right)^{-4} \, 
\label{intensity}
\end{equation}
\begin{eqnarray}
\mbox{\boldmath$\beta$} _\nu & = & \frac{3 \mbox{\boldmath$f$}_\nu}
{2 \, +  \, \sqrt{4 \, - \, 3  \left|  \mbox{\boldmath$f$} _\nu \right| ^2}} \, , \label{beta-f}
\end{eqnarray}
the moments as well as the closure relation, Equation (\ref{closure}), 
are consistent with the assumed radiation field.   
Although the closure relation cannot specify the intensity uniquely,
Equation (\ref{intensity}) has a distinctive feature: the entropy is 
minimum \citep{dubroca99}.    We can obtain the angular
distribution by the Lorentz transform of an isotropic radiation field
\citep{levermore84}.   

We evaluate the flux, $ \mbox{\boldmath$F$} _{\nu,i} $, across a cell
boundary between two adjacent cells by using the reconstructed intensity.
We call the two cells, left (L) and right (R), for later convenience.   
The cell surface is assumed to be normal to the $ i $-th direction, i.e.,
$ \left( \mbox{\boldmath$x$} _{\rm R}  \, - \,  \mbox{\boldmath$x$} _{\rm L} \right) 
\times \mbox{\boldmath$e$} _i \; = \; 0 $.   Then the intensity on the
cell surface is evaluated to be
\begin{equation}
I _\nu ^* (\mbox{\boldmath$n$}) \; = \; 
\left\{
\begin{array}{ll}
I _{\nu, {\rm L}} (\mbox{\boldmath$n$}) &
(\mbox{\boldmath$n$} \cdot \mbox{\boldmath$e$} _i \, \ge \, 0 ) \\ 
I _{\nu, {\rm R}} (\mbox{\boldmath$n$})  &
(\mbox{\boldmath$n$} \cdot \mbox{\boldmath$e$} _i \, < \, 0 ) \\
\end{array} 
\right. \, , \label{upwind}
\end{equation}
where $ I _{\nu, {\rm L}} (\mbox{\boldmath$n$})  $ and
$ I _{\nu, {\rm R}} (\mbox{\boldmath$n$})  $ denote the radiation field
reconstructed in L and R cells, respectively.    Equation (\ref{upwind}) is
based on the fact that photons transmit the surface from L to R when
$ \mbox{\boldmath$n$}\cdot\mbox{\boldmath$e$} _i \, \ge \, 0 $ and
from R to L otherwise.  In other words, the intensity on the upwind side
is identified as that on the surface.    Thus the energy flux,
\begin{equation}
F ^* _{\nu,i} \; = \;\int \left( \mbox{\boldmath$e$}_i \cdot 
\mbox{\boldmath$n$} \right)  \, I ^* _\nu (\mbox{\boldmath$n$}) \,
 d \mbox{\boldmath$n$}  \, ,
\end{equation}
is expected to inherit the \lq \lq upwind\rq \rq nature and  accordingly
to be an alternative to the Godunov-type numerical flux.
Similarly the momentum flux is evaluated to be
\begin{eqnarray}
P _{\nu,ij} ^*  
& = & \frac{1}{c} \, \int \, \left( \mbox{\boldmath$e$}_i \cdot 
\mbox{\boldmath$n$} \right) \, \left( \mbox{\boldmath$e$}_j \cdot 
\mbox{\boldmath$n$} \right) I ^* _\nu (\mbox{\boldmath$n$}) \,d \mbox{\boldmath$n$} \, ,
\end{eqnarray}

The numerical flux, $ F ^* _{\nu,i} $, is expressed by an
explicit function of $ E _{\nu, {\rm L}} $, $ \mbox{\boldmath$f$} _{\nu,{\rm L}} $,
 $ E _{\nu, {\rm R}} $, and $ \mbox{\boldmath$f$} _{\nu,{\rm R}} $.
 For later convenience the numerical flux
 is decomposed into two components,
 \begin{eqnarray}
 F _{\nu,i} ^* & = &  F _{\nu,i,{\rm L}} ^{(+)} \; + \; F _{\nu,i,{\rm R}} ^{(-)} \, , 
 \label{numericalF} \\
 F _{\nu,i,{\rm L}} ^{(+)}  & = & \int _{\mbox{\boldmath$n$}  \cdot \mbox{\boldmath$e$}_i \ge  0} 
 \left( \mbox{\boldmath$e$}_i \cdot 
\mbox{\boldmath$n$} \right)  \, I _{\nu,{\rm L}} (\mbox{\boldmath$n$}) \, 
d \mbox{\boldmath$n$}  \, ,  \\
 F _{\nu,i,{\rm R}} ^{(-)}  & = & \int _{\mbox{\boldmath$n$}  \cdot \mbox{\boldmath$e$}_i <  0} 
 \left( \mbox{\boldmath$e$}_i \cdot 
\mbox{\boldmath$n$} \right)  \, I _{\nu,{\rm R}} (\mbox{\boldmath$n$}) \, 
d \mbox{\boldmath$n$}  \, . 
\end{eqnarray} 
The former is evaluated to be
 \begin{eqnarray}
F _{\nu,i,{\rm L}}  ^{(+)} & \equiv & \int _0 ^{\pi/2} \int _0 ^{2\pi}
I_{\nu,{\rm L}} (\theta,~\varphi) \, \cos \theta \, \sin \theta \,
d\theta~d\varphi \\
& = & 
\left[ \frac{ 3 \, q _{\nu,{\rm L}}   \, + \, 6 \, \beta _{\nu,{\rm L}} ^2 \, 
\cos ^2 \psi _{\nu,{\rm L}}  \, q _{\nu,{\rm L}} ^{-1}
\, - \, \beta _{\nu,{\rm L}} ^4~\cos^4 \psi _{\nu,{\rm L}} q _{\nu,{\rm L}} ^{-3} \, + \, 
8 \beta _{\nu,{\rm L}} 
\cos \psi _{\nu,{\rm L}}}{4 \, (\beta _{\nu,{\rm L}} ^2 \, + \, 3)} \right] \, c E _{\nu,{\rm L}} \, , \\
q _{\nu,{\rm L}}  & = & \left( 1 \, - \, \beta _{\nu,{\rm L}} ^2 \sin ^2 \psi _{\nu,{\rm L}} \right) ^{1/2}  \, , 
\end{eqnarray}
where the angular variables, $ \theta $, $ \varphi $, and $ \psi $,  are 
defined to satisty
\begin{eqnarray}
\left( \mbox{\boldmath$e$} _i \cdot \mbox{\boldmath$n$} \right) & = & \cos \theta \, , \\
\left( \mbox{\boldmath$e$} _j \cdot \mbox{\boldmath$n$} \right) & = & \sin \theta \, 
\cos~\varphi \, , \\
\left( \mbox{\boldmath$e$} _k \cdot \mbox{\boldmath$n$} \right) & = & \sin \theta \, 
\sin~\varphi \, , \\
\left(\mbox{\boldmath$e$} _i \cdot  \mbox{\boldmath$\beta$} _{\rm L} \right)  & = & 
\beta _{\nu,{\rm L}}  \, \cos~\psi _{\nu,{\rm L}} \, .
\end{eqnarray} 
The symbols, $ \mbox{\boldmath$e$} _j $ and
$ \mbox{\boldmath$e$} _k $ denote the unit vectors perpendicular to 
$ \mbox{\boldmath$e$} _i $.   The suffix, L, indicates that the variables are
evaluated in cell L.
We used the computer software, Mathematica, to obtain the above integral.
Similarly we obtain the other half, 
\begin{eqnarray} 
F _{\nu,i,{\rm R}}  ^{(-)} & \equiv & \int _{\pi/2} ^{\pi} \int _0 ^{2\pi}
I_{\nu,{\rm R}} (\theta,~\varphi) \, \cos \theta \, \sin \theta \,
d\theta~d\varphi \\
= &  - & 
\left[ \frac{ 3 \, q _{\nu,{\rm R}}  +  6 \, \beta _{\nu,{\rm R}} ^2
 \cos ^2 \psi _{\nu,{\rm R}}  q _{\nu,{\rm R}} ^{-1}
 - \beta _{\nu,{\rm R}} ^4 \cos^4 \psi _{\nu,{\rm R}} q _{\nu,{\rm R}} ^{-3}  -  
8 \beta _{\nu,{\rm R}} \cos \psi _{\nu,{\rm R}}}
{4 \, (\beta _{\nu,{\rm R}} ^2 \, + \, 3)} \right]  c E _{\nu,{\rm R}} \, .
\end{eqnarray}
Note that $ F _{\nu,i} $ coincides with $ F _{\nu,i,{\rm L}} ^{(+)} \, + \, F _{\nu,i,{\rm R}} ^{(-)} $
since $ f  _\nu \, = \, 4 \beta _\nu \, (3 \, + \, \beta _\nu ^2) ^{-1} $. 

\begin{figure}[h]
\begin{center}
\FigureFile(85mm,80mm){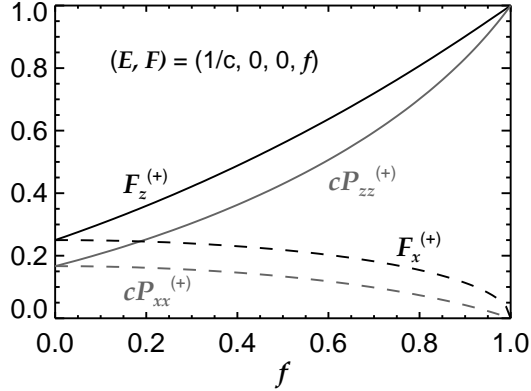}
\end{center}
\caption{Numerical flux obtained by the reconstructed intensity is shown
as a function of $ f $ for $ (E,~\mbox{\boldmath$F$}~=~(1/c,~0,~0,~f)$.  
The black solid and dashed curves denote $ F _z ^{(+)} $ and $ F _x ^{(+)} $,
respectively.  The grey solid and dashed curves denote $ P _{zz} ^{(+)} $
and $ P _{xx} ^{(+)} $, respectively.}\label{numericalFPzz}
\end{figure}

Figure~\ref{numericalFPzz} shows the numerical fluxes evaluated from
the reconstructed intensity for $ (E,~F_x,~F_y,~F _z)~=~(1/c,~0,~0,~f) $.
The black solid and dashed curves denote $ F _z ^{(+)} $ and $ F _x ^{(+)} $ 
as a function of $ f $, respectively.   Both $ F _z ^{(+)} $ and $ F _x ^{(+)} $
approaches to 1/4 in the limit of $ f~=~0 $ (isotropic).   They have the
asymptotic forms of $ F _z ^{(+)} \, \rightarrow \, f $ and 
$ F _x ^{(+)} \, \rightarrow \, \sqrt{(1 \, - \, f)/8} $ in the limit of
$ f \, \rightarrow \, 1 $.

Similarly the radiation pressure is evaluated to be
\begin{eqnarray}
P _{\nu,ii} ^* & = & P _{\nu,ii,{\rm L}} ^{(+)} \; + \;
P _{\nu,ii,{\rm R}} ^{(-)} \, , \label{numericalPii} \\
P _{\nu,ii.{\rm L}} ^{(+)} & \equiv & \int _0 ^{\pi/2} \int _0 ^{2\pi}
I _{\nu,{\rm L}} (\theta,~\varphi) \, \cos^2 \theta \, \sin \theta \,
d\theta~d\varphi \\
& = &  \left[ \frac{ \beta _{\nu,{\rm L}} ^3 \cos ^3 \psi q _{\nu,{\rm L}} ^{-1} \, + \,
3 \beta _{\nu,{\rm L}} \cos \psi _{\nu,{\rm L}} q _{\nu,{\rm L}} \, + \, 4 \beta _{\nu,{\rm L}} ^2 
\cos ^2 \psi _{\nu,{\rm L}} \, + \, 1
\, - \, \beta _{\nu,{\rm L}} ^2}{2 \, (\beta _{\nu,{\rm L}} ^2 \, + \, 3)}  \right] \, , \\
P _{\nu,ii.{\rm R}} ^{(-)} & \equiv & \int _{\pi/2} ^{\pi} \int _0 ^{2\pi}
I_{\nu,{\rm R}}(\theta,~\varphi) \, \cos^2 \theta \, \sin \theta \,
d\theta~d\varphi \\
& = &  \left[ \frac{ - \, \beta _{\nu,{\rm R}} ^3 \cos ^3 \psi q _{\nu,{\rm R}} ^{-1} \, - \,
3 \beta _{\nu,{\rm R}} \cos \psi _{\nu,{\rm R}} q _{\nu,{\rm R}} \, + \, 4 \beta _{\nu,{\rm R}} ^2 
\cos ^2 \psi _{\nu,{\rm R}} \, + \, 1
\, - \, \beta _{\nu,{\rm R}} ^2}{2 \, (\beta _{\nu,{\rm R}} ^2 \, + \, 3)}  \right] \, , \\
P _{\nu,ij} ^{*} & = & P _{\nu,ij,{\rm L}} ^{(+)} \; + \;
P _{\nu,ij,{\rm R}} ^{(-)} \, , \label{numericalPij}  \\
P _{\nu,ij,{\rm L}} ^{(+)}  & \equiv & \int _0 ^{\pi/2} \int _0 ^{2\pi}
I_{\nu,{\rm L}}(\theta,~\varphi) \, \cos \theta \, \sin^2 \theta \, \cos \varphi \,
d\theta~d\varphi \\
& = & \beta _{\nu,j,{\rm L}}\, F _{\nu,i,{\rm L}} ^{(+)} \, , \\
P _{\nu,ij,{\rm R}} ^{(-)}  & \equiv & \int _{\pi/2} ^\pi \int _0 ^{2\pi}
I_{\nu,{\rm R}}(\theta,~\varphi) \, \cos \theta \, \sin^2 \theta \, \cos \varphi \,
d\theta~d\varphi \\
& = & \beta _{\nu,j,{\rm R}} \, F _{\nu,i,{\rm R}} ^{(-)} \, .
\end{eqnarray}
The values of $ P _{zz} ^{(+)} $ and $ P _{xx} ^{(+)} $ are denoted by
the grey solid and dashed curves, respectively, in Figure~\ref{numericalFPzz}.

Remember that our numerical flux is obtained by integrating the intensity over the
left and light hemispheres separately.   \citet{dubroca03} proposed a similar idea
named half space moment approximation.   They integrated the radiative transfer
equation over the half hemisphere and obtained two unknowns and two equations 
for each moment.  Their equations are closed by two closure relations each of 
which is applied for each half hemisphere.  Thus their scheme is different from 
ours in many respects, although the idea, integration over the half hemisphere 
is common.

\subsection{Inclusion of Absorption and Emission within a Cell} 

In the previous subsection we implicitly assumed that the radiation field
is uniform within a cell.    This is not a good approximation when the cell
under consideration itself is optically thick.   The intensity differ 
appreciably on the cell surface from that at the cell center.   Taking
account of the absorption, emission, and scattering within a numerical
cell, we modify Equation (\ref{numericalF}) into
\begin{eqnarray}
F _{\nu,i} ^* & = &  \eta _{\nu,i,{\rm L}}  \left[ \zeta _{\nu,i,{\rm L} }\, 
F _{\nu,i,{\rm L}} ^{(+)} \, +  \, \left( 1 \, - \, \zeta _{\nu,i,{\rm L}} \right) \,
\frac{c E _{\nu,{\rm L}}}{4} \right]
\; + \;  \left(1 \, -\, \eta _{\nu,i,{\rm L}} \right) \,
\pi \,  B _{\nu,{\rm L}}  \nonumber \\
& \; & + \; \eta _{\nu,i,{\rm R}}  \left[ \zeta _{\nu,i,{\rm R} }\, 
F _{\nu,i,{\rm R}} ^{(+)} \, -  \, \left( 1 \, - \, \zeta _{\nu,i,{\rm R}} \right) \,
\frac{c E _{\nu,{\rm R}}}{4} \right]
\; - \;  \left(1 \, -\, \eta _{\nu,i,{\rm R}} \right) \,
\pi \,  B _{\nu,{\rm R}}  \; , \label{numericalF2}  \\
\eta _{\nu,i,{\rm L}} & = & \exp~\left[ -  \, \frac{\kappa _{\nu,a,{\rm L}} \, \rho \, 
\Delta x _{i,{\rm L}}}{2} \right] \, , \\
\eta _{\nu,i,{\rm R}} & = & \exp~\left[ -  \, \frac{\kappa _{\nu,a,{\rm R}} \, \rho \,
\Delta x _{i,{\rm R}}}{2} \right] \, , \\
\zeta_{\nu,i,{\rm L}} & = & \exp~\left[ -  \, \frac{\kappa _{\nu,s,{\rm L}} \,  \rho \,
\Delta x _{i,{\rm L}}}{2} \right] \, , \\
\zeta _{\nu,i,{\rm R}} & = & \exp~\left[ -  \, \frac{\kappa _{\nu,s,{\rm R}} \, \rho \,
\Delta x _{i,{\rm R}}}{2} \right] \, .
\end{eqnarray}
Equation (\ref{numericalF2}) denotes an approximation to the formal soultion
of the radiative transfer in the limit of $ \Delta x \, \rightarrow \, 0 $.  
On the other hand, in the limit of of  $ \Delta x \, \rightarrow \,
\infty $, it describes the state in which the radiation is 
in the thermal equilibrium in each cell.

Similarly Equations (\ref{numericalPii}) and (\ref{numericalPij}) are modified into
\begin{eqnarray}
P _{\nu,ii} ^* & = &  \eta_{\nu,i,{\rm L}}  \left[ \zeta _{\nu,i,{\rm L} }\, 
P _{\nu,ii,{\rm L}} ^{(+)} \, +  \, \left( 1 \, - \, \zeta _{\nu,i,{\rm L}} \right) \,
\frac{E _{\nu,{\rm L}}}{6} \right]
\; + \;  \left(1 \, -\, \eta _{\nu,i,{\rm L}} \right) \,
\frac{2\pi}{3c} \,  B _{\nu,{\rm L}}  \nonumber \\
& \; & + \; \eta _{\nu,i,{\rm R}}  \left[ \zeta _{\nu,i,{\rm R} }\, 
P _{\nu,ii,{\rm R}} ^{(-)} \, +  \, \left( 1 \, - \, \zeta _{\nu,i,{\rm R}} \right) \,
\frac{E _{\nu,{\rm R}}}{6} \right]
\; + \;  \left(1 \, -\, \eta _{\nu,i,{\rm R}} \right) \,
\frac{2\pi}{3c} \,  B _{\nu,{\rm R}}  \; ,   \label{numericalPii2} \\
P _{\nu,ij} ^* & = & \eta  _{\nu,i,{\rm L}} \zeta _{\nu,i,{\rm L}} \, P _{\nu,ij,{\rm L}} ^{(+)} 
\; + \; \eta  _{\nu,i,{\rm R}} \zeta _{\nu,i,{\rm R}} \, P _{\nu,ij,{\rm R}} ^{(-)} 
\label{numericalPij2}
\end{eqnarray}

We prove that our modified numerical flux is asymptotic preserving and reproduces
the diffusion approximation in the limit of 
$ \kappa _a \rho \Delta x~\rightarrow~\infty $.   For later convenience, 
we define the symbol
\begin{equation}
\Delta F_{\nu}  ^* \; \equiv \;  \sum _i \frac{1}{\Delta x _i} \, \left[
F _{\nu,i} ^* \left(\mbox{\boldmath$x$} \, + \, \frac{\Delta x _i}{2} \, 
\mbox{\boldmath$e$} _i  \right) \, - \,
F _{\nu,i} ^* \left(\mbox{\boldmath$x$} \, - \, \frac{\Delta x _i}{2} \, 
\mbox{\boldmath$e$} _i  \right)  \, \right] ,
\end{equation}
to denote the \lq \lq divergence'' of  the numerical flux evaluated on the cell.
Similary we define 
\begin{equation}
\Delta P_{\nu,ij}  ^* \; \equiv \;  \sum _k  \frac{1}{\Delta x _k} \, \left[ 
P _{\nu,ij} ^* \left(\mbox{\boldmath$x$} \, + \, \frac{\Delta x _k}{2} \, 
\mbox{\boldmath$e$} _k  \right) \, - \,
P _{\nu,ij} ^* \left(\mbox{\boldmath$x$} \, - \, \frac{\Delta x _k}{2} \, 
\mbox{\boldmath$e$} _k  \right)  \right] \, ,
\end{equation}
to denote the divergence of the pressure tensor.
The discretized M1 equations reduce to
\begin{eqnarray}
\Delta F _{\nu} ^* & = & - \kappa _{\nu} \, \rho \,
\left( c E _\nu \, - \, 4 \pi B _\nu \right)  , \label{EqF} \\
c  ^2 \sum _i \Delta P _{\nu,ij} ^* & = & - \,
\left( \kappa _{\nu,a} \, + \, \kappa _{\nu,s} \right) \, \rho \, c \,  F _{\nu,j} .
\label{EqP}
\end{eqnarray}
in radiative equilibrium.

When $ \Delta x $ is large and the temperature difference is small,
Equations (\ref{numericalF2}), (\ref{numericalPii2}), and
(\ref{numericalPij2}) reduce to 
\begin{eqnarray}
F _{\nu,j} ^* & \simeq & \pi \, \left[ B _{\nu} \left( T _{\rm L}  \right) 
\, - \, B _{\nu} \left( T _{\rm R} \right) \right] \\
& \simeq & \pi \, \frac{\partial B _{\nu}}{\partial T} \, 
\left( T _{\rm L} \, - \, T _{\rm R} \right) , \label{diffF} \\
P _{\nu,ii} ^* & \simeq & \frac{2\pi}{3}  \left[ B _{\nu} \left( T _{\rm L} \right) 
\, +  \, B _{\nu} \left( T _{\rm R} \right) \right] \, \label{diffPjj} , \\
P _{\nu,ij} ^* & \simeq & 0  \hskip 0.5cm {\rm if} \; i \, \ne \, j , \label{diffPij}
\end{eqnarray}
respectively.   

Substituting  Equations (\ref{diffF}) through (\ref{diffPij}) into Equations
(\ref{EqF}) and (\ref{EqP}), we obtain
\begin{eqnarray}
E _\nu & \simeq & \frac{4\pi}{c} \, B _\nu \\
F _{\nu,j} & \simeq & \frac{4 \pi}{3 \rho} \, \frac{1}{\kappa _{\nu,a} \, + \, \kappa _{\nu,s}}
\, \frac{\partial B _\nu}{\partial T} \, 
\frac{\partial T}{\partial x _j}  \, ,
\end{eqnarray}
where only the most dominant terms are taken into account for simplicity.
These equations are equivalent to the diffusion approximation
(see, e.g, \cite{castor04}, for frequency dependent diffusion
approximation).  Thus  our numerical flux gives a good approximation 
in the optically thick limit.    We can expect that our numerical flux gives
a reasonable approximation at any optical depth.

\subsection{Second Order Accuracy in Space}

The numerical flux given in the previous subsection is of the first order accuracy
in space.    A numerical flux of the second order accuracy in space can be 
obtained by  applying  Monotone Upwind-centered Scheme for Conservation Laws 
(MUSCL, see e.g., \cite{hirsch90}).  MUSCL evaluates the physical states on the cell
boundaries from the left and right hand sides by extrapolation.  MUSCL applies
a limiter such as minmod function in order to avoid spurious extrema from
the values obtained by simple extrapolation.   We need some additional  cares
to avoid unphysical extrapolation when applying MUSCL to M1 equations.
Otherwise, the energy density can be smaller than the radiative energy flux
divided by the speed of light ($ E \, < \, |\mbox{\boldmath$F$}/c| $).   We 
use the formulae given in Appendix  to achieve the second order
accuracy in space.

The numerical flux given in Appendix is constructed on the assumption
that both emission and scattering change linearly along the line from
the cell center to the boundary.    The emission and scattering on the
cell boundary are evaluated by linear extrapolation with the minmod
limiter, i.e., by MUSCL.   Therefore the contribution of the emission 
and scattering within the cell is expressed as the linear combination
of those evaluated at the cell center and boundary.   Further details
are given in Appendix.

\subsection{Time Evolution}

Using the notation given in the previous subsection, we can rewire the
M1 equations in the form,
\begin{eqnarray}
\frac{\partial E _\nu}{\partial t} & = &   
\kappa _{\nu,a} \rho \left( c E _\nu \, - \, 4 \pi B _\nu \right) \,  - \,
\Delta F _\nu \, ,   \label{timeE} \\
\frac{\partial F _{\nu,j}}{\partial t} & = & - \,
\left( \kappa _{\nu,a} \, + \, \kappa _{\nu,s} \right) \, \rho c \, F _{\nu,j} 
\; -  \;  c ^2 \Delta P _{\nu,j} \; . \label{timeF}
\end{eqnarray}
We use two different methods to integrate Equations (\ref{timeE}) and (\ref{timeF}) 
in the first order accuracy in time 
depending on the problem.   We use the forward difference of the first
order in time in flash test in which we solve propagation of radiation from
a sphere into vacuum.  In the rest of test problems, 
we use a  formal solution for forwarding the energy density and flux in time.   

The formal solution of of  Equations (\ref{timeE}) and (\ref{timeF})  expressed as 
\begin{eqnarray}
E _\nu (t + \Delta t) & =& e ^{- \kappa _{\nu,a} \rho c \Delta t} \, E _\nu (t) 
\; + \; \left( 1 \, - \, e ^{- \kappa _{\nu,a} \rho c \Delta t} \right) \,
\left( B _\nu \, + \, \frac{\Delta F _\nu}{\kappa _{\nu,a} \rho} \right) \, , 
\label{formalE}  \\
F _{\nu,j} (t + \Delta t) & =  & e ^{- (\kappa _{\nu,a} + \kappa _{\nu,s})  \rho  c \Delta t} \, 
F _{\nu,j} (t) \; + \; \left[ 1 \, - \,    e ^{- (\kappa _{\nu,a} + \kappa _{\nu,s})  \rho c \Delta t} 
\right]  \, \frac{c ^2 \Delta P _{\nu,j}}{(\kappa _{\nu,a} + \kappa _{\nu,s}) \rho}
\label{formalF} \, ,
\end{eqnarray}
where all the physical variables are evaluated at time, $ t $.  Thus our scheme
is explicit in the sense $ E _\nu $ and $ F _{\nu,j} $ are obtained without 
iteration.   Nevertheless, the formal solution allows to use a much
longer time step than the simple forward difference in time 
when $ \kappa _{\nu,a} \rho \Delta x _j \, \gg \, 1 $.   The time step should
be smaller than $ \Delta t \, \le \, 1/(\kappa _{\nu,a} \rho) $ in the simple
forward difference.   When $ \kappa _{\nu,a} \rho \Delta x \, \gg \, 1 $,
the constraint is serious since it is much shorter than the time for
propagation of signal, $ \Delta s/ c $.   

We integrate the hydrodynamical equations by an explicit scheme.
The time step is taken to be the minimum of the thermal and hydrodynamical
timescales.

A solution of the second order accuracy in time can be obtained by a two step
Runge-Kutta method.   We used the average of time derivatives evaluated at
$ t~=~t _0 $ and $ t _0 \, + \, \Delta t $ when integrating equations from
$ t~=~t _0 $ to $ t _0 \, + \, \Delta t $.

\section{Monochromatic Test Problems}

\subsection{Shadow Test}

We performed a shadow test to illustrate the effects of including absorption
within the cell.    We exposed uniform monochromatic radiation to 
a square absorber of $ \kappa \rho~=~50 $.   Scattering and emission
are neglected for simplicity.   The computation box 
covers the square of $ 0 \, \le \, x \, \le \, 12 $ and $ 0 \, \le \, y \, \le \, 6 $.
The absorber occupies the square region of $ 2 \, \le \, x \, \le 3 $ 
and $ 0 \, \le \, y \, \le \, 2 $.   The spatial resolution is 
$ \Delta x~=~\Delta y~=~0.1 $.  We imposed the uniform radiation,
$ (E,~F_x,~F_y)~=~(1.0,~0.999~c,~0) $, from $ x~=~0.0 $.   Reflection
boundary is applied to the upper and lower boundaries of $ y~=~6$
and 0.   The outgoing boundary is placed on the right boundary,
$ x~=~12.0 $, so that no radiation enters from the boundary.  
We realized the outgoing boundary by vanishing the flux from 
outside the boundary.   It is a virtue of the kinetic flux
that the outgoing boundary is easily constructed.  
We obtained the
equilibrium state by solving the time dependent M1 model with
the time step, $ \Delta t~=~0.5~\Delta x/c $.  

\begin{figure}[h]
\begin{center}
\FigureFile(120mm,80mm){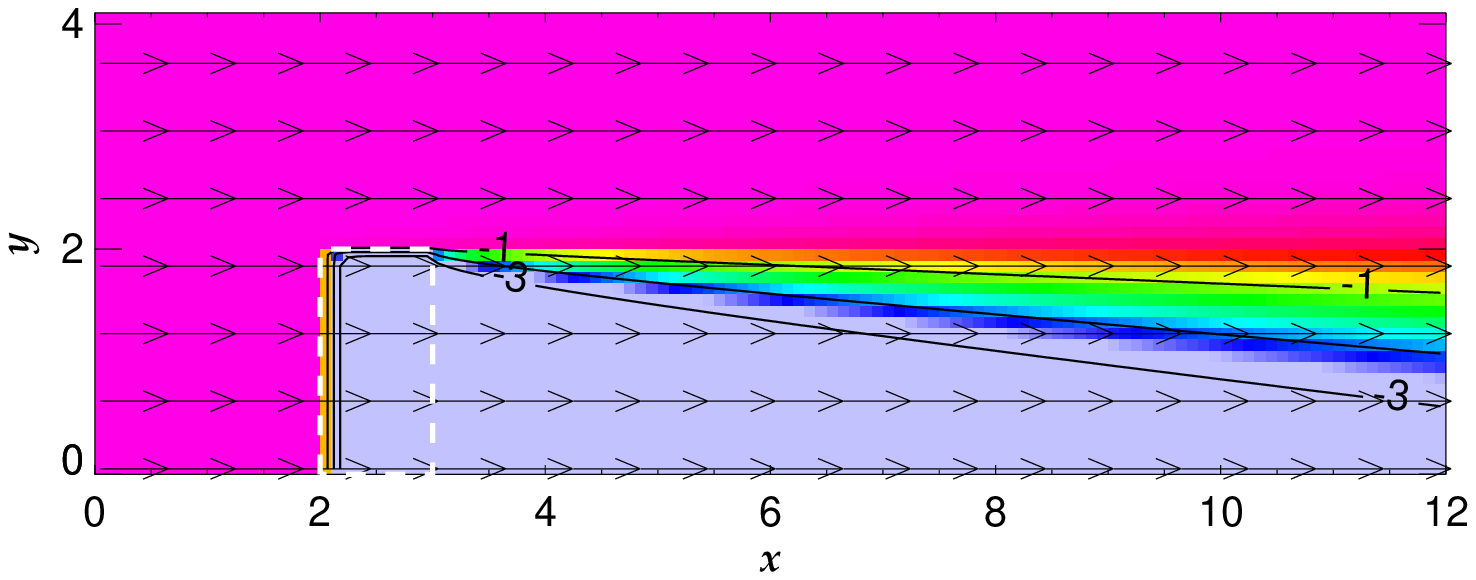}
\FigureFile(120mm,80mm){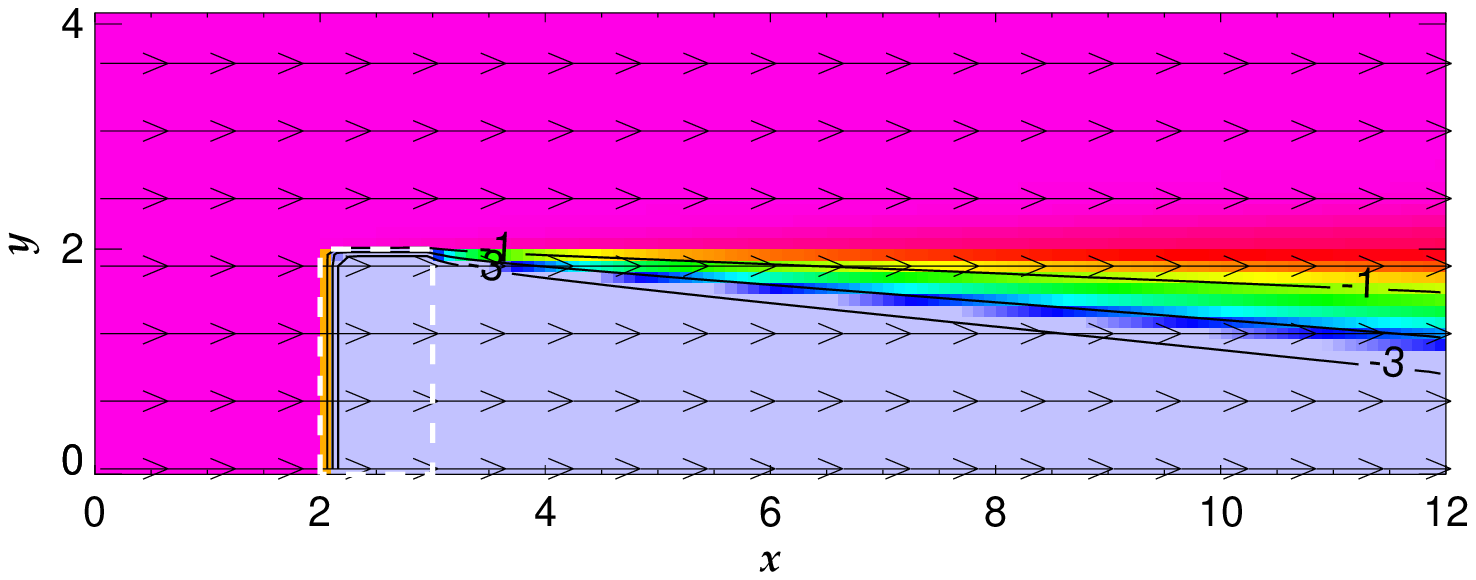}
\end{center}
\caption{2 Dimensional Shadow Test.    The upper and lower panels show
results obtained by the numerical flux of the first order accuracy in space and
that of the second order, respectively.    The dashed line denotes the
absorber.   The brightness denotes the  radiation energy density in 
the logarithmic scale.   The solid curves denote the contours of
$ \log E~=~-3,~2,~{\rm and}~-1 $.   See text for more details.
The arrows denote the vector, 
$ \mbox{\boldmath$F$}/|\mbox{\boldmath$E$}|$}\label{shadow}
\end{figure}

Figure~\ref{shadow} shows the result of the shadow test in the
equilibrium state.   The numerical flux of the first order accuracy
in space is used in the upper panel while that of the second order
accuracy is used in the lower panel.
The brightness denotes the energy density
in logarithmic scale.  Note that the energy density drops very 
sharply behind the left side of the absorber in the time step.   
The arrows denote
$ \mbox{\boldmath$F$}/(c E _\nu)$.  

Very weak radiation shines in from the upper right corner behind 
the absorber because the incident radiation is not perfectly 
beamed ($ F/(cE)~=~0.999 $).    FWHM of the beam is
$ \theta _{\rm FWHM}/2~=~1.^\circ 58 $, since it is
evaluated to be $ \theta _{\rm FWHM}/2~\simeq~0.870\,\sqrt{1\,-\,f} $ from Equations
(\ref{intensity}) and (\ref{beta-f}) for $ 1 \, - \, \beta \, \simeq \, 2 (1 \, - \, f) \, \ll \, 1 $.   
 The intensity decreases by a factor 100 at $ \theta~=~5.^\circ 33 $.   
 The inclination of the contour of  $ \log~E~=~-2 $ is $ 6^\circ $ and $ 5^\circ $ from
 the horizontal line in the solutions of the first and second order accuracy,
 respectively and constant with the intensity distribution of $ f~=~0.999 $.
 The solutions are different only in dark area of $ \log~E~<~-2 $. 
 The contour of $ \log~E~=~-3 $ is more inclined in the solution of the
 first order accuracy than in that of the second order accuracy.   
  
\subsection{Flash Test}

We performed the following test for studying the propagation of radiation
in vacuum.   The initial radiation field at $ t~=~0 $ is set to be
\begin{equation}
\left(E,~F_x,~F_y,~F_z \right) \; = \; 
\left\{
\begin{array}{ll}
(1,~c f,~0,~0) & {\rm if} \; |\mbox{\boldmath$r$} | \, \le \, r _0  \\
(0,~0,~0,~0) & {\rm if} \; |\mbox{\boldmath$r$} | \, > \, r _0 
\end{array}
\right. \, ,
\end{equation}
so that uniform radiation is confined within the sphere of  $ r \, \le \,  r _0 $.
No absorption and emission are assumed in this test.   Then the radiation is expected
to  be confined in the spherical shell of $ r _0 \, - \, ct \, \le \, r \, \, \le \, r _0 \, + \, c t $.

\begin{figure}[h]
\begin{center}
\FigureFile(85mm,85mm){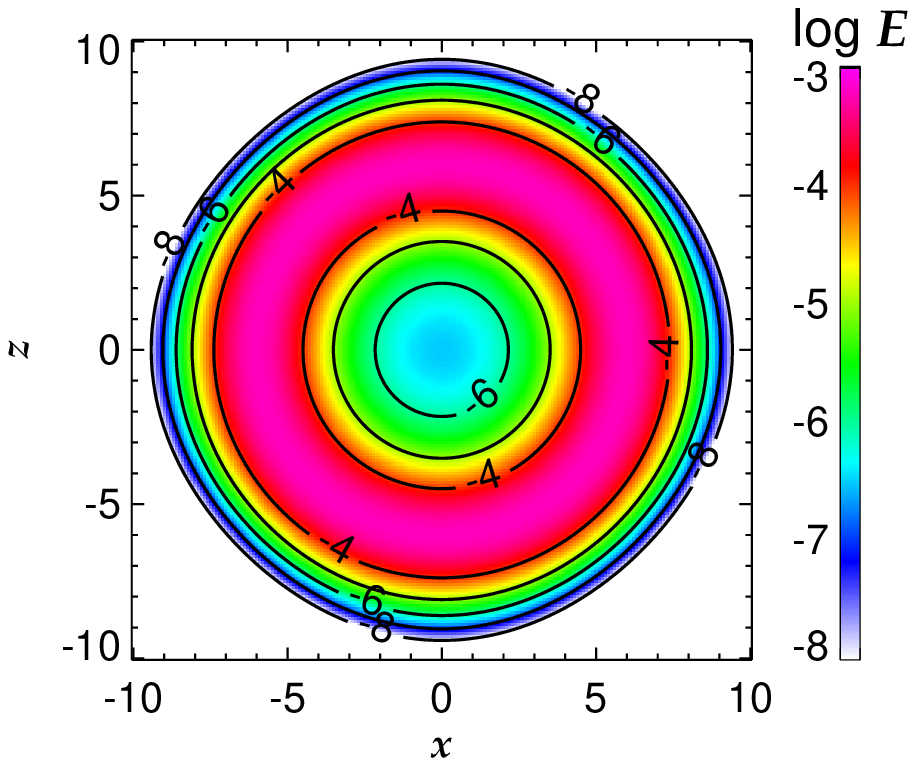} \\
\FigureFile(85mm,85mm){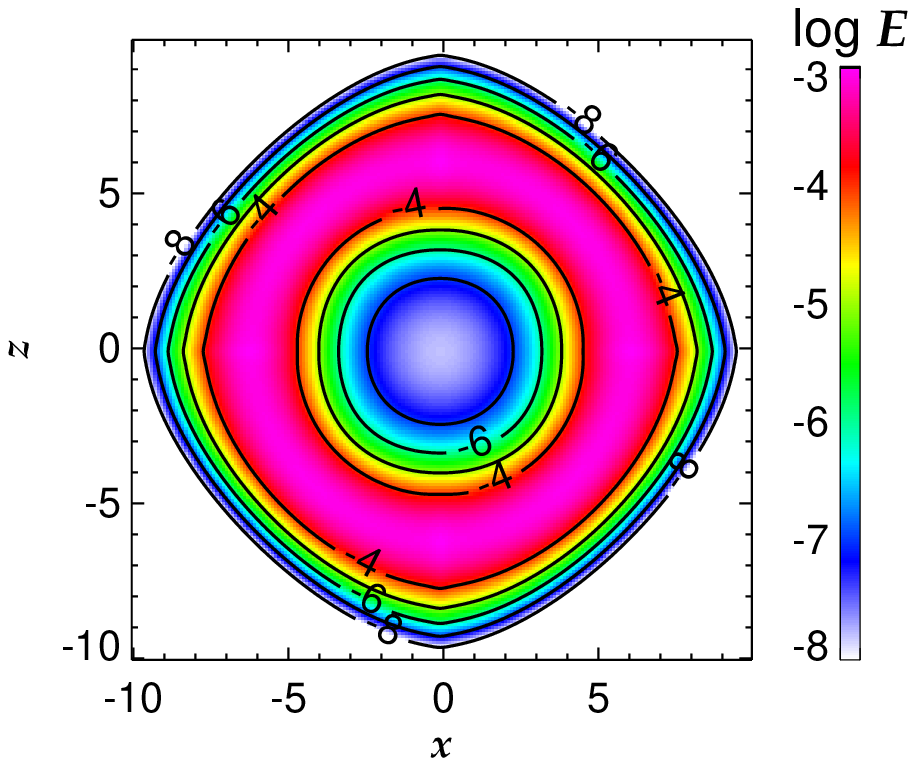}
\end{center}
\caption{The result of flash test for $ f = 0 $.  The upper panel
shows the solution by HLLE scheme and the lower panel does
that by our numerical flux.   Both the panels denote the solutions of
the first order accuracy in space and in time.} \label{flash0}
\end{figure}

Figure~\ref{flash0} denotes the radiation energy density for $ f~=~0.0 $ and
$ r _0~=~0.5 $ at $ t~=~6.0/c $.   The spatial resolution and time step are taken
to be $ \Delta x = 0.1 $ and $ \Delta t~=~0.3~\Delta x/c $, respectively.   The upper panel
denotes the result obtained by the simple HLLE flux, while the lower panel
does that obtained with our numerical flux.  Both of them denote the solutions
of the first order accuracy in space and in time.
The brightness and the contours 
denote $ \log E $.   It should be noted that HLLE does not work at
$ \Delta t~=~0.35~\Delta x/c $, while our method does 
at $ \Delta t~=~0.5~\Delta x /c $.   A longer time step can be taken
safely if we apply the kinetically reconstructed numerical flux.

If we solve the flash problem perfectly by taking the full angular dependence,
the radiation should be confined an expanding spherical shell of 
$ c t \, - \, r _0~\le~r~\le~ct \, + \, r _0 $.   It should be isotropic around 
the origin and the flux should be radial.     

Our model  shows a higher contrast  than the simple HLLE model.  
However a clear anisotropy of numerical origin appears in our model.
This anisotropy is due to the fact that the change in the flux direction in 
a cell is not taken into account.

\begin{figure}[h]
\begin{center}
\FigureFile(85mm,85mm){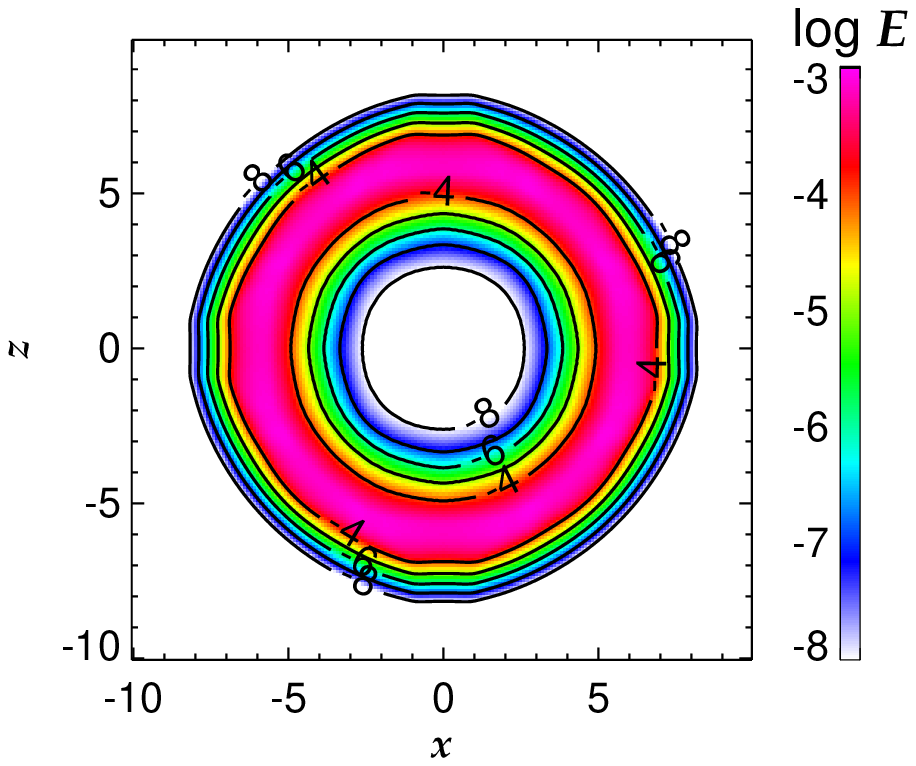} \\
\FigureFile(85mm,85mm){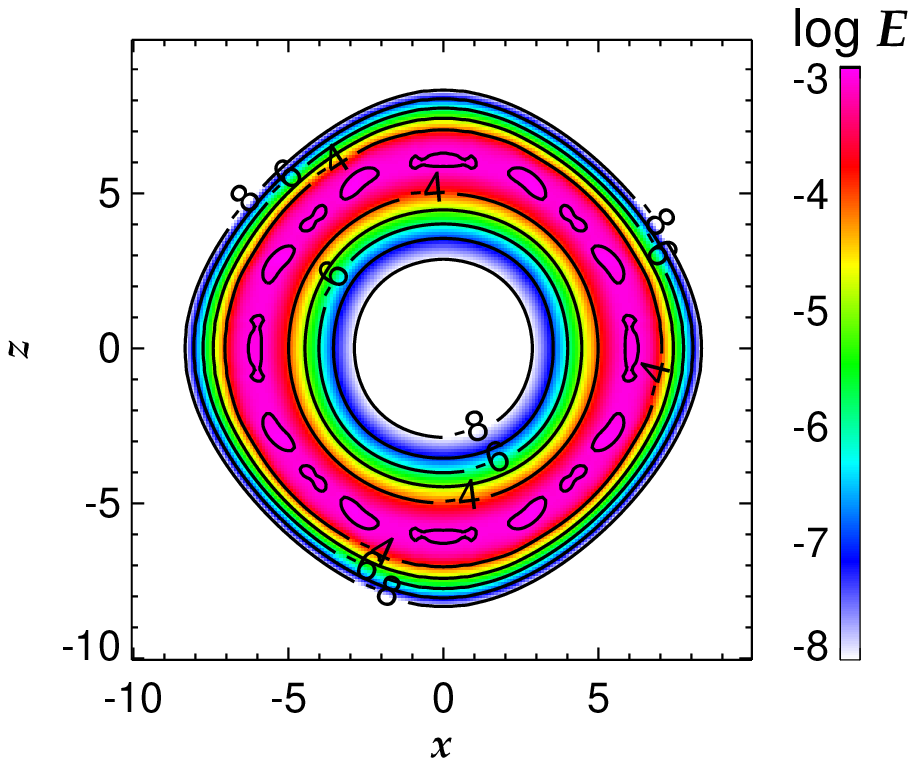}
\end{center}
\caption{The same as Fig.~\ref{flash0} but for the solutions of the 
second order accuracy in space and in time} \label{flash0-2nd}
\end{figure}

Figure~\ref{flash0-2nd} is the same as Figure~\ref{flash0} but for the
solutions of the second order accuracy in space and in time.    The 
anisotropy has been removed in the solutions of the second order
accuracy.    The low contrast of the simple HLLE  model is also
improved, although the contrast is still higher in our model.

\begin{figure}[h]
\begin{center}
\FigureFile(80mm,70mm){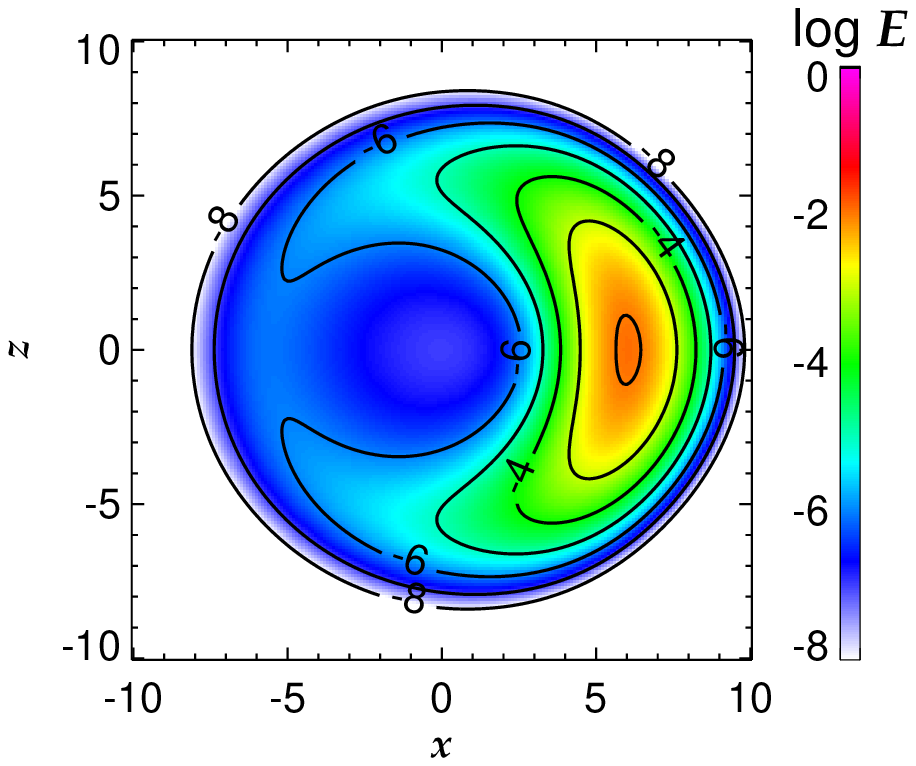} \\
\FigureFile(80mm,70mm){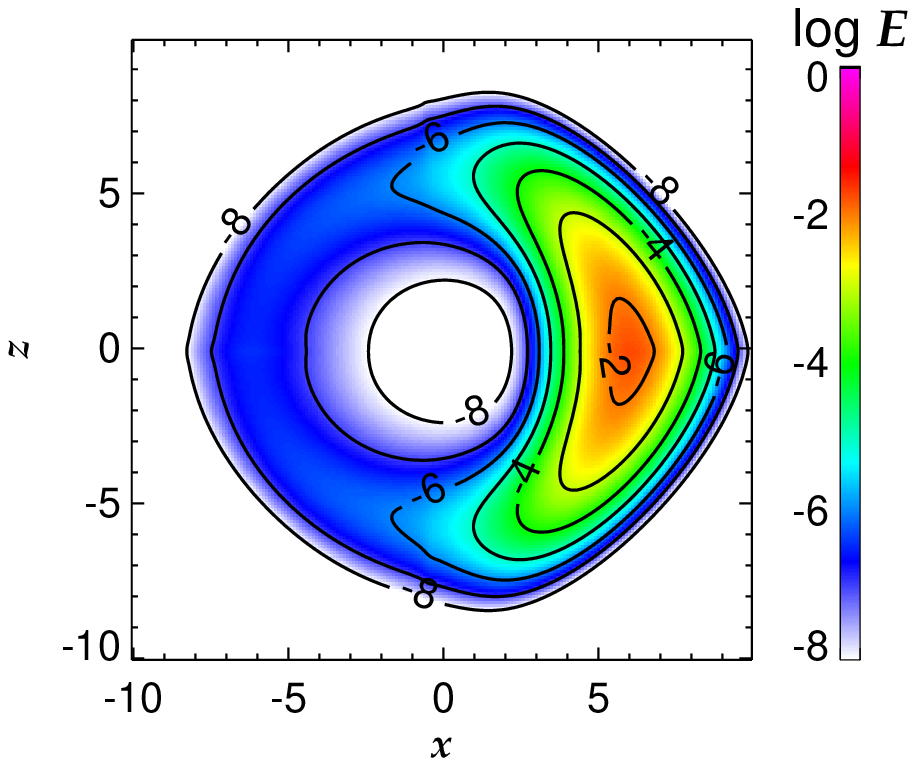}
\end{center}
\caption{The same as Fig.~\ref{flash0} but for $f = 0.9$.}\label{flash09}
\end{figure}

Figure~\ref{flash09} is the same as Figure~\ref{flash0} but for $ f~=~0.9 $.
When $ f $ is close to 1.0, the forward radiation is much stronger than the
backward one.    However, the shell of the high radiation energy density
should expand spherically as in case of $ f~=~0$, if we take the full 
angular dependence of the radiation.  It expands as expected in the 
lower panel (kinetically reconstructed
numerical flux), while the center of the sphere shows a spurious shift in the
upper panel (HLLE).   This is due to the characteristic speeds of waves in
M1 model equations.    Their absolute values are smaller than the speed of
light and all the characteristic speeds have the same sign 
when $ f \, > \, 0.69 $ (see e.g., Figure 1 of \cite{gonzalez07} for the 
characteristic speeds as a function of $ f $).   Thus the center of radiation 
shell shifts rightward in the HLLE solution.

One might think that the shift would be the nature of M1 model equations 
and should be reproduced in numerical simulations.   However, we 
should realize that the characteristic speed changes according to that
in $ f ~=~F/(cE)$.    The ratio, $ f $, decreases on the left edge of the radiation
sphere, since radiation moving rightward has a higher $ f $ than the 
initial value.  Once it decreases to the critical value on the edge, 
the radiation begins  to propagates leftward and it approaches to $ f~=~-1 $.
It should take only an instant  for the left edge of the radiation 
sphere to propagate at the speed of $ - c$, if our spatial resolution is 
extremely high.    However,
it takes a few time steps for $ f $ to decrease to the critical value in
a cell close to the left edge.   The propagation to rightward is delayed
a few time steps in the HLLE model.   

\begin{figure}[h]
\begin{center}
\FigureFile(80mm,70mm){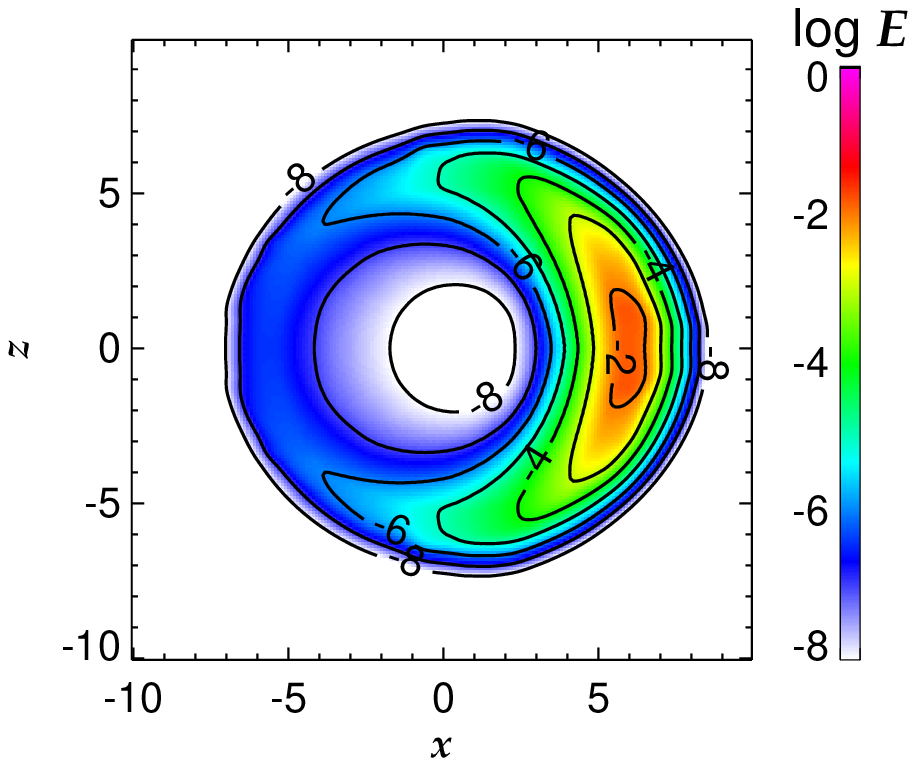} \\
\FigureFile(80mm,70mm){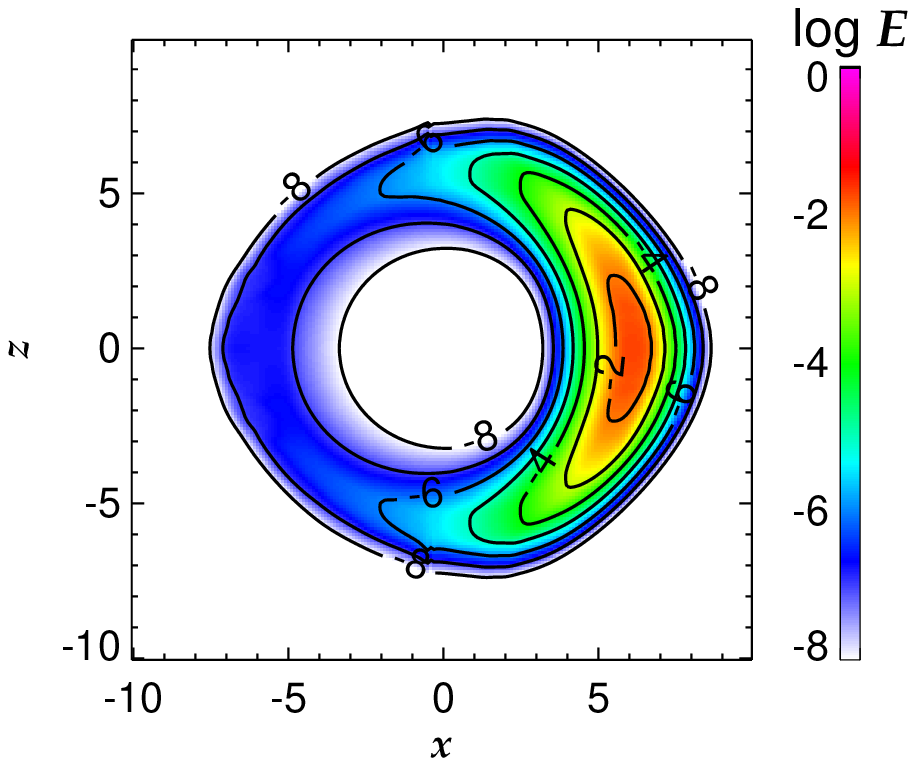}
\end{center}
\caption{The same as Fig.~\ref{flash09} but for 
the solutions of second order accuracy in space and in time.}\label{flash09-2nd}
\end{figure}

Figure~\ref{flash09-2nd} shows the second order accurate solutions for
the flash test of $ f~=~0.9 $.    The bright shell is more sharply captured 
in both the second order accurate solutions.   The contrast is still higher
in our model than in the HLLE model.    The central dark hole is shifted
leftward in the HLLE model.   

\section{Application to Irradiated Protoplanetary Disks}

Young stars are often associated with gaseous disks called protoplanetary
disks (see, e.g, a review by \cite{williams11}).   
They are irradiated by the radiation from the central 
stars to shine at various wavelengths.    They reflect optical
 and near infrared stellar lights and emit mid and far infrared ones.   
Both absorption and scattering are dominated by dusts, although
the optical properties remain somewhat unknown.   It is essential to
take account of the frequency dependent opacity when modeling
protoplanetary disks.   Optical and near infrared radiation 
heat up the disks from outside while mid and far infrared emission 
cool down them from inside.    The mid and far infrared flux from inside 
balances the optical and near infrared one from outside in equilibrium.

\begin{figure}[h]
\begin{center}
\FigureFile(85mm,80mm){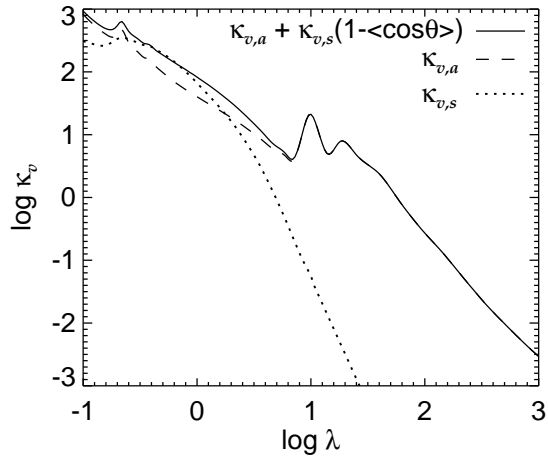}
\end{center}
\caption{The opacity used in our application to irradiated protoplanetary
disks.   The curves denote, $ \kappa _{\nu,a} $, $ \kappa _{\nu,s} $,
and  $ \kappa _{\nu,a} \, + \, \kappa _{\nu,s} \, (1 \, - \, \left\langle
\cos \theta \right\rangle ) $ as a function of the wavelength, 
$ \lambda \, = \, c / \nu $.}\label{opacity}
\end{figure}

We apply our numerical method to an irradiated protoplanetary disk
as a test for multi-color problem.   We use the opacity table of
\citet{draine03} in which $ \kappa _{\nu,a} $, $ \kappa _{\nu,s} $,
and $ \langle \cos \theta \rangle $ are given as a function of the
wavelength.   Applying the spline fit to the
table, we obtained the values at the wavelengths,
\begin{equation}
\log \left( \frac{\lambda _m}{1.0~\mu {\rm m}} \right)  \; = \; 0.02~m \, ,
\label{lambda} 
\end{equation}
where $ m $ is a integer in the range of $ -50~\le~m~\le~150 $.
Figure~\ref{opacity} shows $ \kappa _{\nu,a} $, $ \kappa _{\nu,s} $,
and $ \kappa _{\nu,a} \, + \, \kappa _{\nu,s} (1 \, - \, \langle \cos \theta \rangle)  $
as a function of $ \lambda $.   The opacity is obtained under the 
assumption that the dust occupies  1\% of the total mass.  In other
words, we did not take account of sedimentation of dust for simplicity.

 We use two sets of M1 model; one denotes the direct
radiation from the central star and the other does scattered by and 
reemitted from the protoplanetary disks.  Then  radiation energy
density and flux are expressed as
\begin{eqnarray}
E _\nu & = & E _\nu ^\prime \, + \, E _\nu ^{\prime\prime} \, , \\
\mbox{\boldmath$F$} _\nu & = & \mbox{\boldmath$F$} _\nu ^\prime
\, +  \, \mbox{\boldmath$F$} _\nu ^{\prime\prime}  \, ,
\end{eqnarray}
where the symbols with prime denote the values of the direct
stellar radiation and those with double prime do the total of
scattered radiation and emission from the disk.  We apply the
closure relation separately for the two components.   The M1 model
equations are expressed as
 \begin{eqnarray}
 \frac{\partial E _\nu ^\prime }{\partial t} \; + \;
\sum _{i=1} ^3
 \frac{\partial F _{\nu,i} ^\prime}{\partial x _i}  & =  &  - 
 \left[ \kappa _{\nu,a} \, +\, \kappa _{\nu,s} \, (1 \, - \, \langle \cos \theta \rangle 
 \right]  \, \rho \, c  E _\nu ^\prime  \, .  \\
  \frac{\partial E _\nu ^{\prime\prime} }{\partial t} \; + \;
\sum _{i=1} ^3
 \frac{\partial F _{\nu,i} ^{\prime \prime}}{\partial x _i}  & =  &  
 \left[ \kappa _{\nu,s} \, (1 \, - \, \langle \cos \theta \rangle )
 \right]  \, \rho \, c E _\nu ^\prime  \, - \;
\kappa _{\nu,a} \, \rho \, \left[ c E  _\nu ^{\prime\prime}  \, - \,
4 \pi B _\nu (T) \right] \, ,  \\
\frac{\partial F _{\nu,i}  ^{\prime}}{\partial t} \; + \;
c ^2 \sum _{j=1} ^3 P _{\nu,ij} ^{\prime} & = & - \, 
\left[ \kappa _{\nu,a} \, + \, \kappa _{\nu,s} \, 
(1 \, - \, \langle \cos \theta \rangle \right]  c F _{\nu,i} ^\prime \, , \\
\frac{\partial F _{\nu,i}  ^{\prime\prime}}{\partial t} \; + \;
c ^2 \sum _{j=1} ^3 P _{\nu,ij} ^{\prime\prime} & = & - \, \left[ \kappa _{\nu,a} \, + \, \kappa _{\nu,s} \, 
(1 \, - \, \langle \cos \theta \rangle)  \right]  c F _{\nu,i} ^{\prime\prime}  \, .
 \end{eqnarray}
 The separation of the direct stellar light from the rest radiation avoids spurious beam
 collision on the disk surface.  
 The radiation energy densities are evaluated at the wavelengths given by Equation 
 (\ref{lambda}), i.e., 201 bands in the range of 
 $ 0.1~\mu{\rm m}~\le~\lambda~\le~1~{\rm mm} $.   Thus our model has the 
 spectral resolution of  $ \Delta \lambda~=~4.61\times 10^{-2}~\lambda $.
 This spectral resolution is good enough to study the spectral features of dust
 opacity.
 
 We solved the above moment equation of radiation and the hydrodynamical equations
 simultaneously.  In this paper we restrict ourselves to the disk in equilibrium in which
 the emission from the disk balances the heating by irradiation.   We ignored self
 gravity of the disk and viscous heating by accretion for simplicity.   The radial 
 component of the gravity is assumed to be balanced with the centrifugal force
 due to the disk rotation.
 
 We assume that the central star has the mass and radius, $ M _* $ and 
 $ R _* $, respectively.  The stellar radiation is assumed to be the black
 body of $ T  _{\rm eff} $.     We made both one and two dimensional models
 of the protoplanetary disk with the numerical flux of the first order accuracy
 in space.  They are described  in subsections  4.1 and 4.2,
 respectively. 
 
\subsection{1D Model Based on the Grazing Recipe} 

Our 1D model describes the vertical structure of the protoplanetary disk
at a given radius, $ R $,  from the central star.  The stellar radiation, 
$ E  _\nu ^\prime $, is evaluated to be
\begin{eqnarray}
E _\nu ^\prime (r,~z) & = & \frac{\pi}{c} \, \left( \frac{R_*}{R} \right) ^2 \, B _\nu (T _{\rm eff} )\,
\exp \, \left[ - \, \frac{\tau _\nu(z)}{\alpha}  \right] \, , \\
\tau _\nu (z)  & = & \left[ \kappa _{\nu,a} \, + \, \kappa _{\nu,s} 
\, \left( 1 \, - \, \langle \cos \theta \rangle \right) \right]  \, 
\int _z ^\infty \rho (r,~z^\prime) \, dz^\prime \, , \\
\alpha & = & \left[ 0.4 \, \frac{R _*}{R} \, + \, R \, \frac{d}{d R} \left( 
\frac{H _*}{R} \right)  \right] \, ,
\end{eqnarray}
according to the grazing angle recipe \citep{chiang97}.
Here $ H _* $ denotes the height of the \lq photosphere' at which the
stellar radiation is attenuated by a factor of $ e ^{-1} $.  We evaluated
the photosphere at $ \lambda~=~0.302~\mu{\rm m} $.  We evaluated
$ d(H_*/R)/dR $ consistently by solving the vertical structure at a slightly
different radii, $ 0.891~R $ and $1.122~R$.  

We obtained the steady state solution by integrating the M1 model
equations and equations for hydrostatic balance simultaneously.
We used the Lagrangian coordinate in this 1D model.   

\begin{figure}[h]
\begin{center}
\FigureFile(80mm,80mm){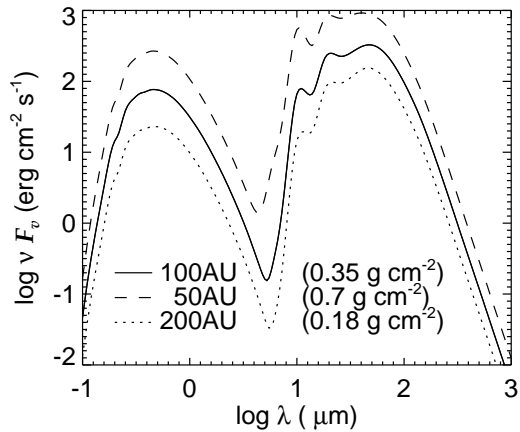}
\end{center}
\caption{Model spectra for a protoplanetary disk irradiated by the star
of $ M~=~2.0~M_\odot $, $ R~=~2.5~R _* $, and $ T_{\rm eff}$~=~9500K.
Each curve denotes the flux from the disk surface at the designated radius.}
\label{1Dspectra}
\end{figure}

Figure~\ref{1Dspectra} shows the model spectra at $ R~=~50$,
100, and 200 AU from the star of $ M _*~=~2.0~M_\odot $, 
$ R _*~=~2.5~R _\odot $, and $ T _{\rm eff} $~=~9,500~K.
These parameters are taken to be similar to those of AB Aur \citep{vandenacker97} 
for which inner hole and spirals structure are seen in optical \citep{grady99}
and near infrared  (\cite{fukagawa04,hashimoto11}).

The surface density is assumed to be
\begin{equation}
\Sigma \;  = \; 0.35 \; \left( \frac{R}{100~{\rm AU}} \right) ^{-1} \;
{\rm g~cm}^{-2} \, .
\end{equation}
The model spectra are consistent with those obtained by 
\citet{dalessio98,dullemond01},
who solved the angle dependent radiative transfer with the Monte
Carlo simulation.   The ratio of the grazing angle to the disk aspect
is $  d/d \ln R [\ln (H_*/R)]~=~0.251 $ at $ R~=~100~{\rm AU} $,
which is close to the standard value, $ 2/7~=~0.282 $ (\cite{chiang97}).     
The value depends rather on the opacity and surface density distribution.
The standard value was obtained by a simplified analytical model.  
The difference is not numerical.  

\begin{figure}[h]
\begin{center}
\FigureFile(80mm,80mm){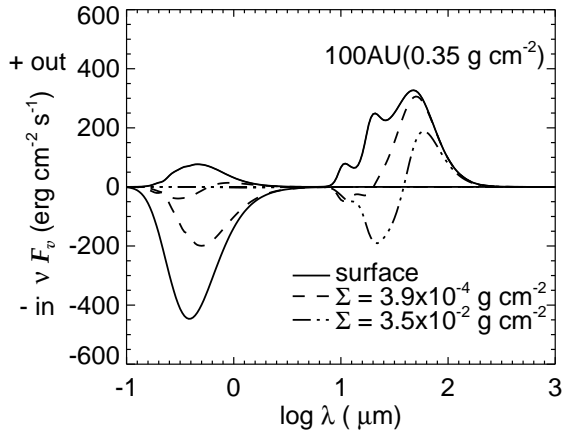}
\end{center}
\caption{The wavelength dependent energy flux.  The surface density
denotes that above the layer.}\label{1Dflux}
\end{figure}

Figure~\ref{1Dflux} confirms that the heating is balanced with the cooling
correctly in our solution.   Each curve denotes the energy flux,  
$ \nu F _{\nu,z} ^\prime $ and $ \nu F _{\nu,z}  ^{\prime\prime} $ at a 
given height as a  function of the wavelength.   The value is taken to be 
positive when the energy flows outward from the disk.  The solid curves
denote $ F _{\nu,z} ^\prime $ (negative) and $ F _{\nu,z} ^{\prime\prime} $
(positive) at the surface (not the photosphere).   The dashed curve
denotes $ F _{\nu,z} ^{\prime\prime} $ at the level above which the disk
has the surface density, $ 3.9 \times 10 ^{-4} $ g~cm$^{-2}$. 
The dash dotted curve does that at $ \Sigma $~=~$ 3.5 \times 10 ^{-2} $
g~cm$^{-2}$.   It is clearly shown that the net flux vanishes at any hight.
The mid infrared is the main heating source in the layer below 
$ \Sigma \, > \, 3.5 \times 10 ^{-2} $ gm~cm$^{-2}$.

\subsection{2D Axisymmetric Model}

1D model assumes implicitly that the surface density changes 
gradually with the radius.   However some protoplanetary disks
may have holes and the surface density may rise sharply at some
radius.   The transition disks are thought to be the case (see, e.g., a review
by \cite{williams11} and the references therein).
We construct a two dimensional model  of a transitional disk assuming
the symmetry around the the axis and that on the mid plane. 

Following \citet{honda12},  who made spectral energy distribution
(SED) model for a transitionl disk around  HD169142,
we made a model in which the surface density
is expressed as
\begin{equation}
\ln \Sigma \; = \; \frac{1}{2} \, \left\{  
\ln \left( \Sigma _{\rm in} \Sigma _{\rm out} \right) \; + \; 
\ln \left( \frac{\Sigma _{\rm out}}{\Sigma _{\rm in}} \right) \, 
\, \tanh \left[ \Gamma \, \ln \left( \frac{r}{r _0} \right)  \right]  \right\} \; - \;
\ln \left( \frac{r}{r _0} \right) \, ,
\end{equation}
where $ \Sigma _{\rm in} $, $ \Sigma _{\rm out} $, $ \Gamma $, and
$ r _0 $ are model parameters to be chosen.  We obtained the 
radiation and gas in the region of $ 40~{\rm AU} \, \le \, r \; \le  \, 160~{\rm AU} $
and $ |z| \, \le \, 80~{\rm AU} $, 
with the resolution of $ \Delta r~=~0.3~{\rm AU} $ and $ \Delta z~=~0.4~{\rm AU} $
in the cylindrical coordinates. 
We placed the reflection boundary on $ Z~=~0 $ and the outgoing boundary 
on $ z~=~80~{\rm AU} $.   The incoming flux from $ R $~=~40~AU was fixed.
The flux incoming from $ r $~=~160~AU was assumed to balance with the 
outgoing one in the disk and to vanish outside the disk.  
The mass, radius, and effective temperature of the central star are
assumed to be $ M~=~2.0~M_\odot $,  $ R _*~=~1.25 \times 10 ^{-2}$~AU,
and $ T _{\rm eff}~=~9,000~$K, respectively.

\begin{figure}[h]
\begin{center}
\FigureFile(85mm,85mm){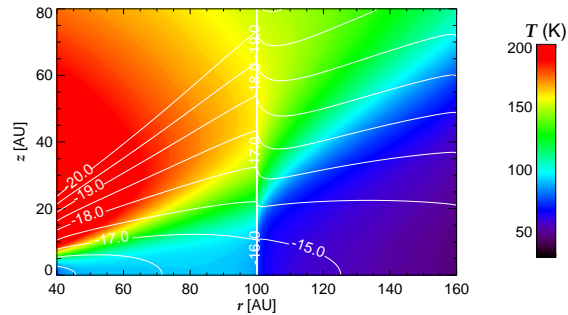}
\end{center}
\caption{The temperature and density distribution in the model having the
sharp rise in the surface density at $ r~=~100$~AU.   The greenness denotes
the temperature and the curves denote the contours of $ \log \rho $ in the
interval of $ \Delta \log \rho~=~0.5$.} \label{2Dsharp_rhoT}
\end{figure}

Figure~\ref{2Dsharp_rhoT} shows the density and temperature distribution
in equilibrium for $ r _0 $~=~100~AU, $ \Gamma~=~\infty $,
$ \Sigma _{\rm in} $~=~$3.5\times 10^{-3}$~g~cm$^{-3}$, and 
$ \Sigma _{\rm out} $~=~$3.5\times 10^{-1}$ ~g~cm$^{-2}$.    The wall at $ r~=~100~{\rm AU} $
is heated up to $ T~\simeq 140~{\rm K} $ and hence expands more.  
The grayness denotes the temperature and the curves denote the contours
of $ \log \rho $ in the interval of $ \Delta \log \rho~=~0.5$.  Note
that we solved the vertical hydrodynamic balance consistently while the
vertical density distribution is given in \citet{honda12}.   The radiation from
the wall heats up the inner disk by 15~K in the range of 70~AU$ \, \le \, r \, \le \,$
100~AU compared with the model without the wall, i.e.,
$ \Sigma \, = \, \Sigma _{\rm in} \, \left( r / r _0 \right) ^{-1} $.

\begin{figure}[h]
\begin{center}
\FigureFile(85mm,85mm){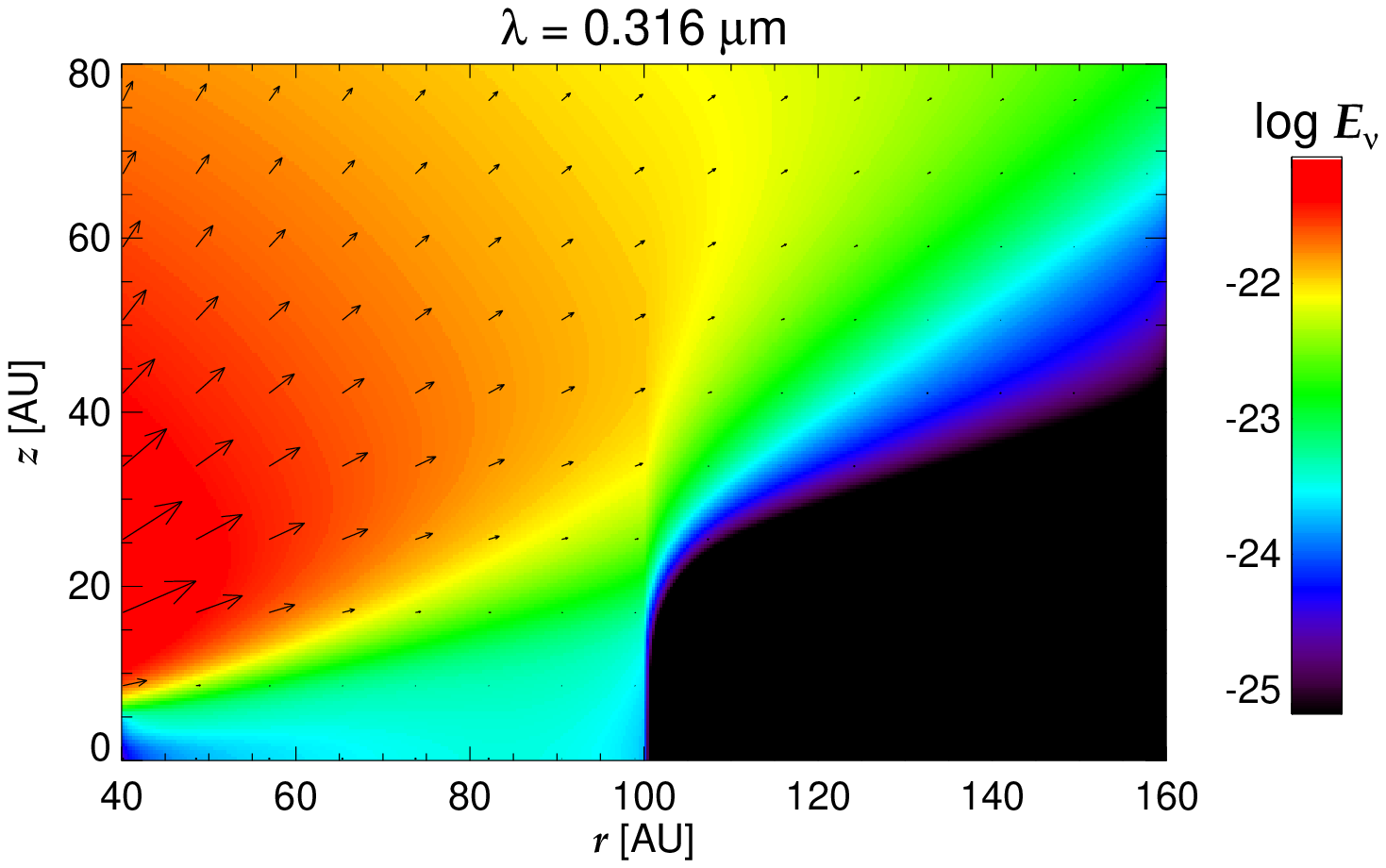} \\
\FigureFile(85mm,85mm){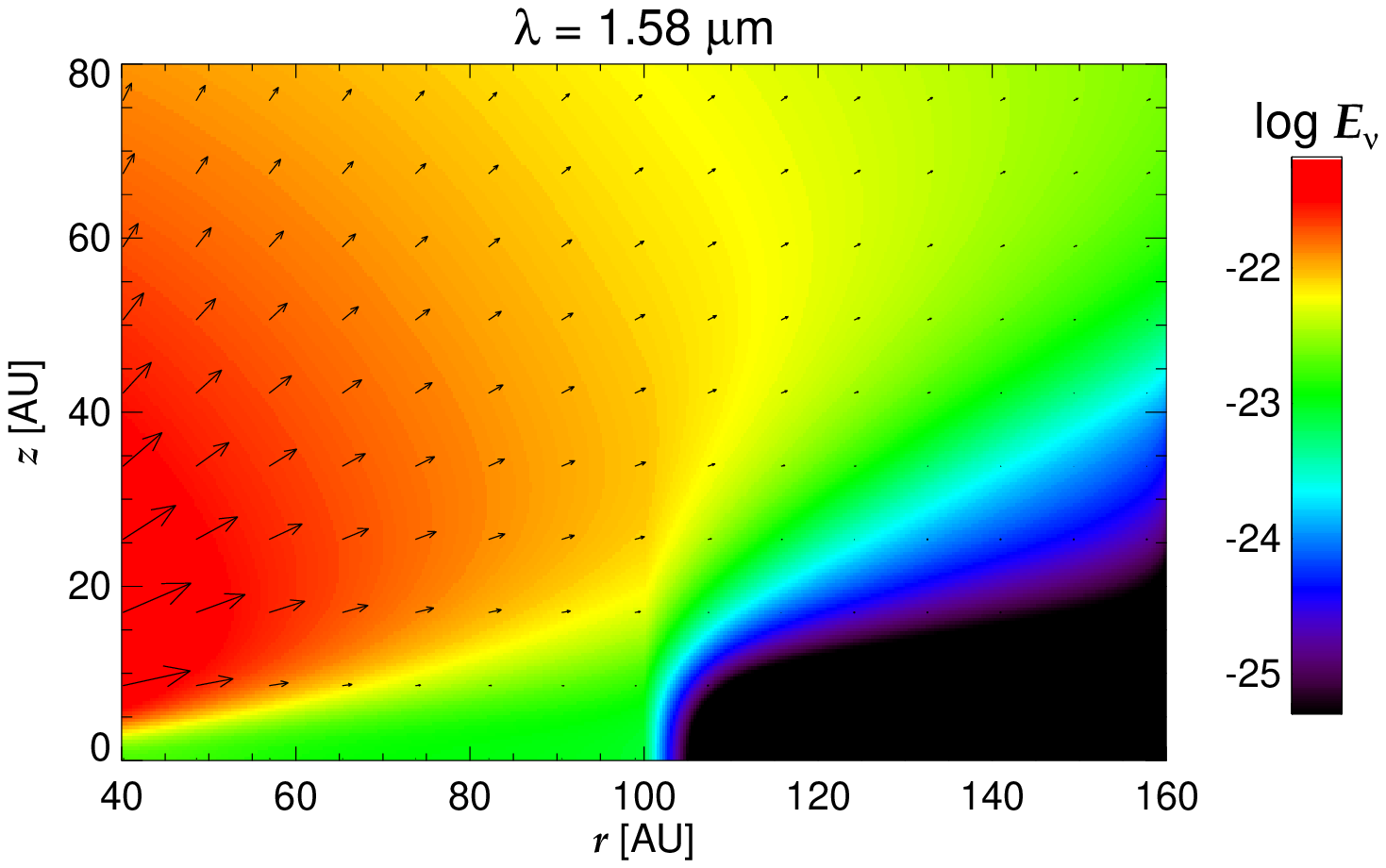} \\
\FigureFile(85mm,85mm){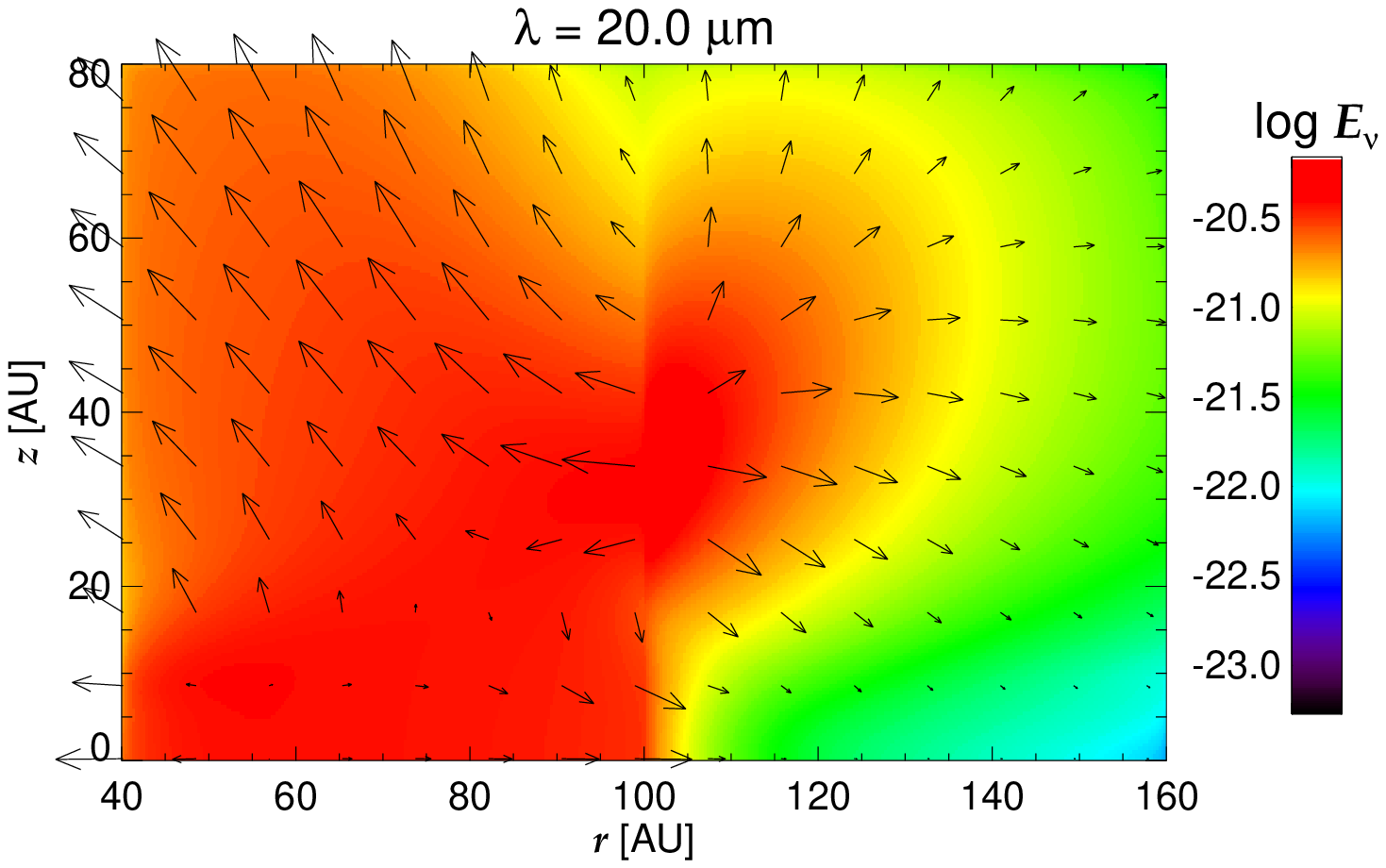} \\
\FigureFile(85mm,85mm){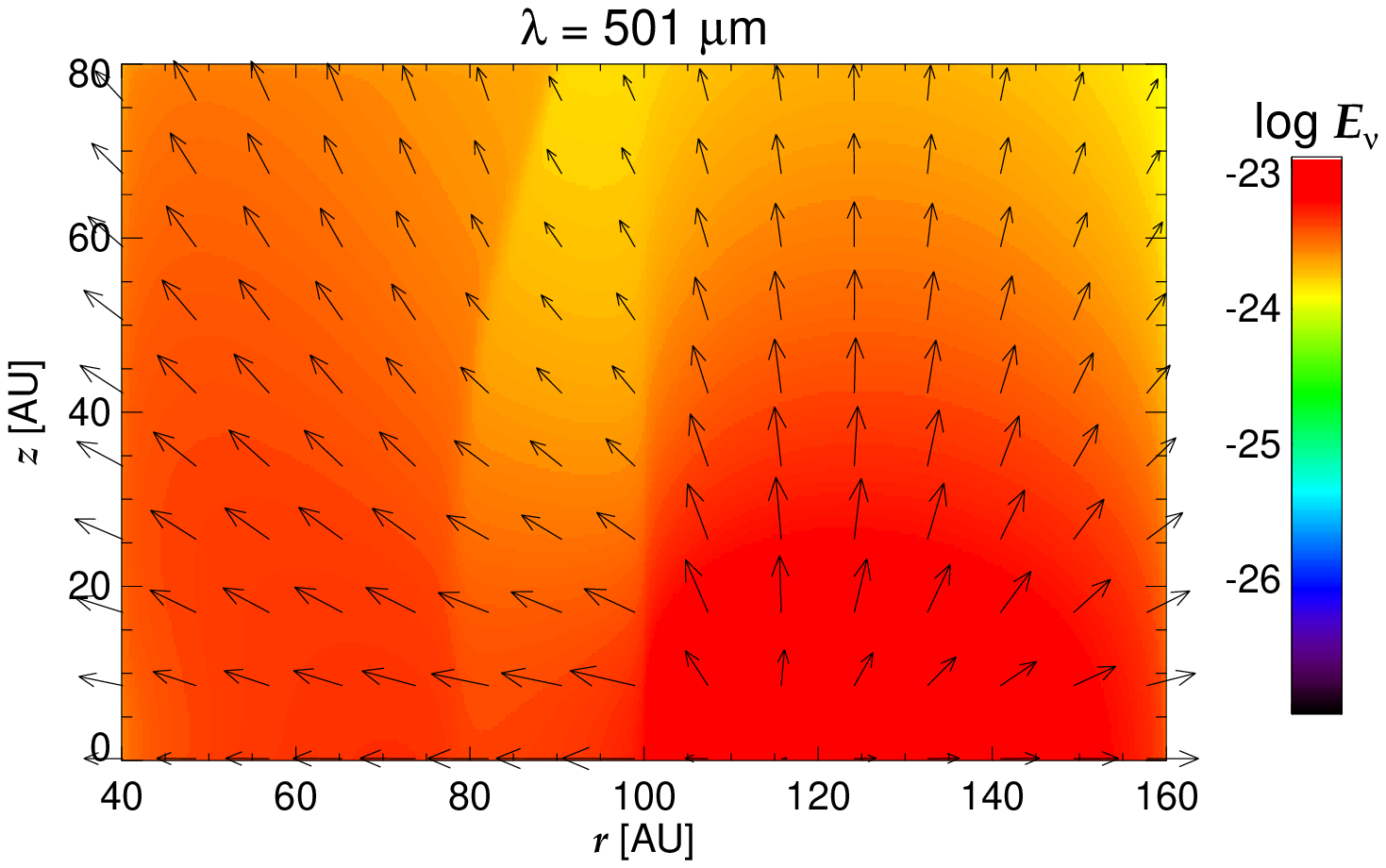}
\end{center}
\caption{Total energy distribution, $ E _\nu (r,~z) $ at  0.316~$\mu$m, 
1.58~$\mu$m (H-band),  and
20~$\mu$m (Q-band), and 501~$\mu$m from top.}\label{Erad_m0}
\end{figure}

Figure~\ref{Erad_m0} shows the total energy density, $ E _\nu $, at
$ \lambda $~= 0.316~$\mu$m, 1.58~$\mu$m,  
20~$\mu $m, and 501~$\mu$m from top to bottom.  The color denotes
$ \log E _\nu $ in unit of $ {\rm erg~cm}^{-3}~{\rm Hz} ^{-1} $ as indicated
in the left bar.  The arrows denote the vector
$ \mbox{\boldmath$F$} _\nu / E _\nu $.   

At $ \lambda~=~0.316~\mu{\rm m} $,
stellar radiation  absorbed or scattered on an upper layer of low density
and does not penetrate into protoplanetary disks.   Near infrared 
radiation penetrate a little deeper interior of the disk but does not reach
the mid plane.  The upper part of the wall is bright  at $ \lambda~=~20~\mu$m.
The wall is heated up by  stellar light and emits mid and far infrared radiation.   
 The disk is more transparent at a longer wavelength as shown in the bottom 
 panel of Figure~\ref{Erad_m0}.   The direction of energy flux also depends
 on the wavelength.   The energy flows from the central star in the optical
 and near infrared, while it flows from the wall and upper surface in the
 mid infrared and from the disk in the far infrared.
 
\begin{figure}[h]
\begin{center}
\FigureFile(80mm,85mm){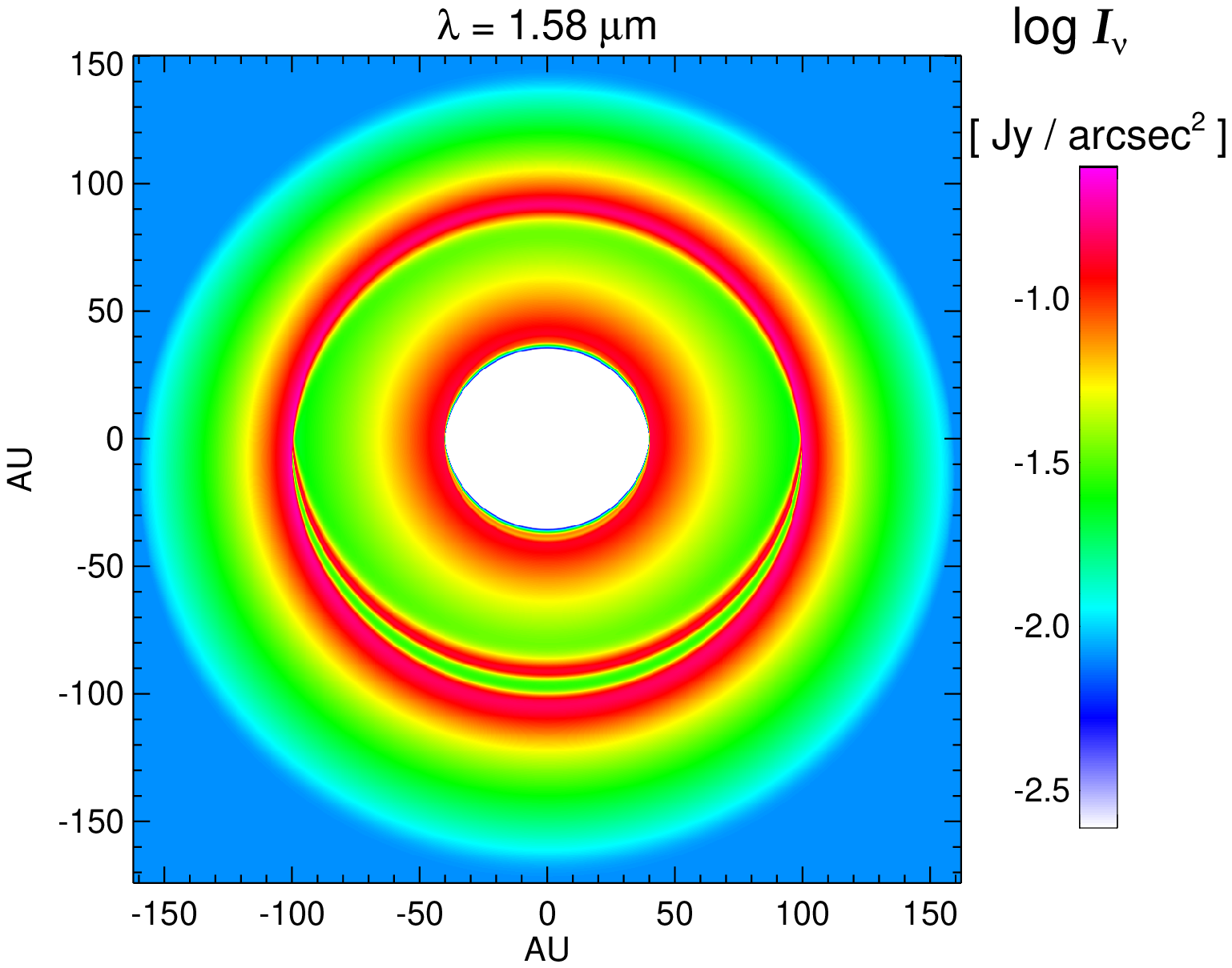}
\FigureFile(80mm,85mm){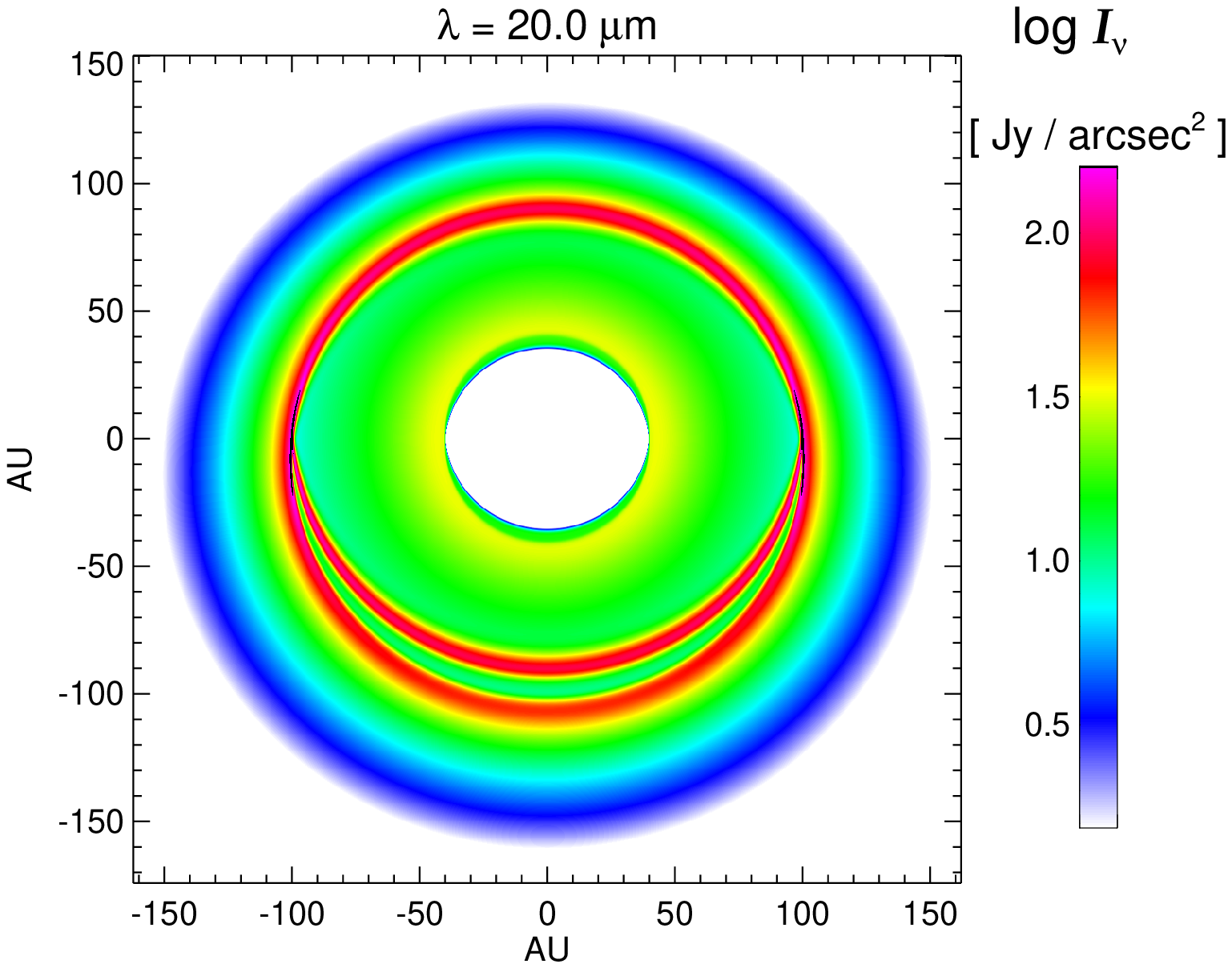} \\
\FigureFile(80mm,85mm){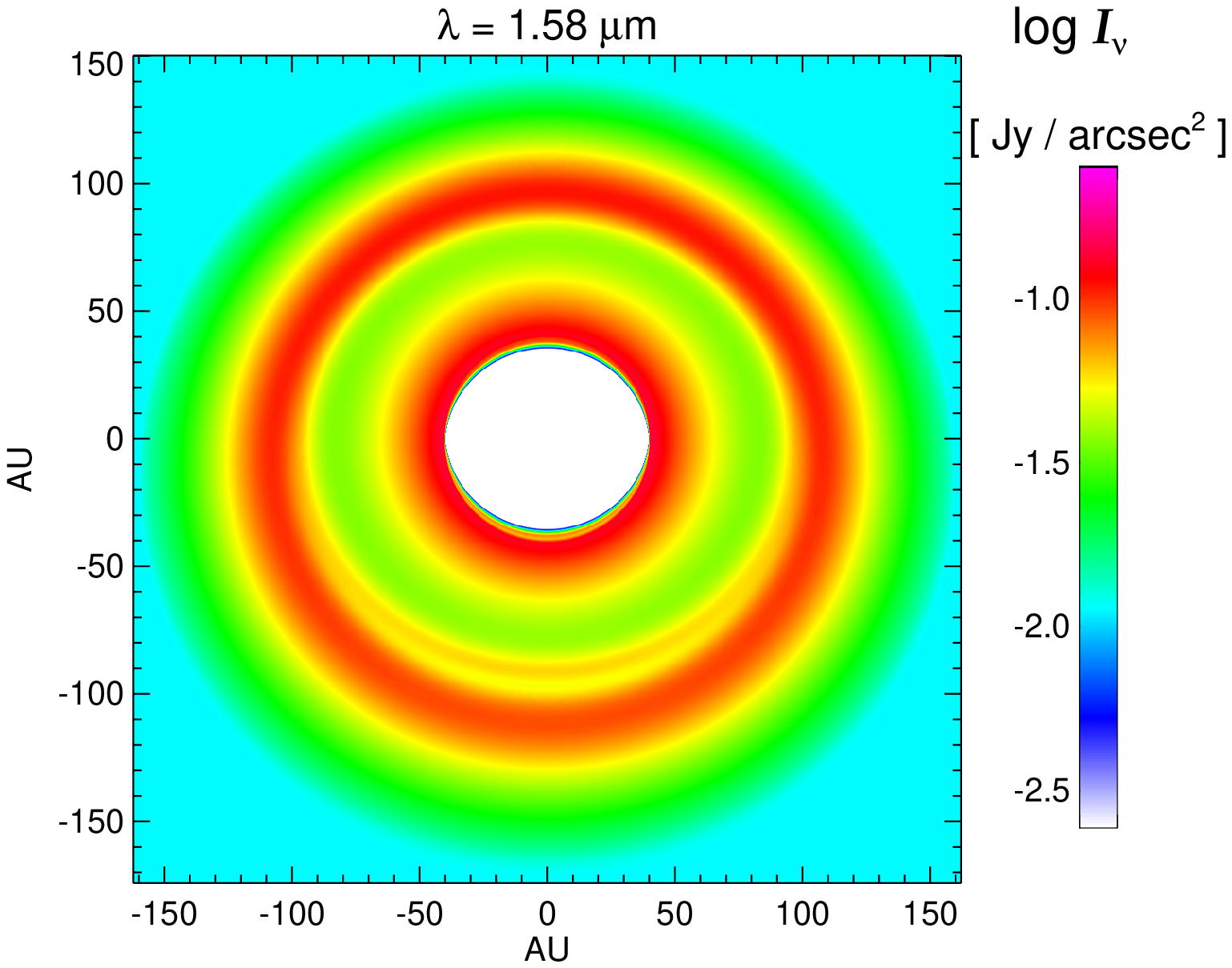}
\FigureFile(80mm,85mm){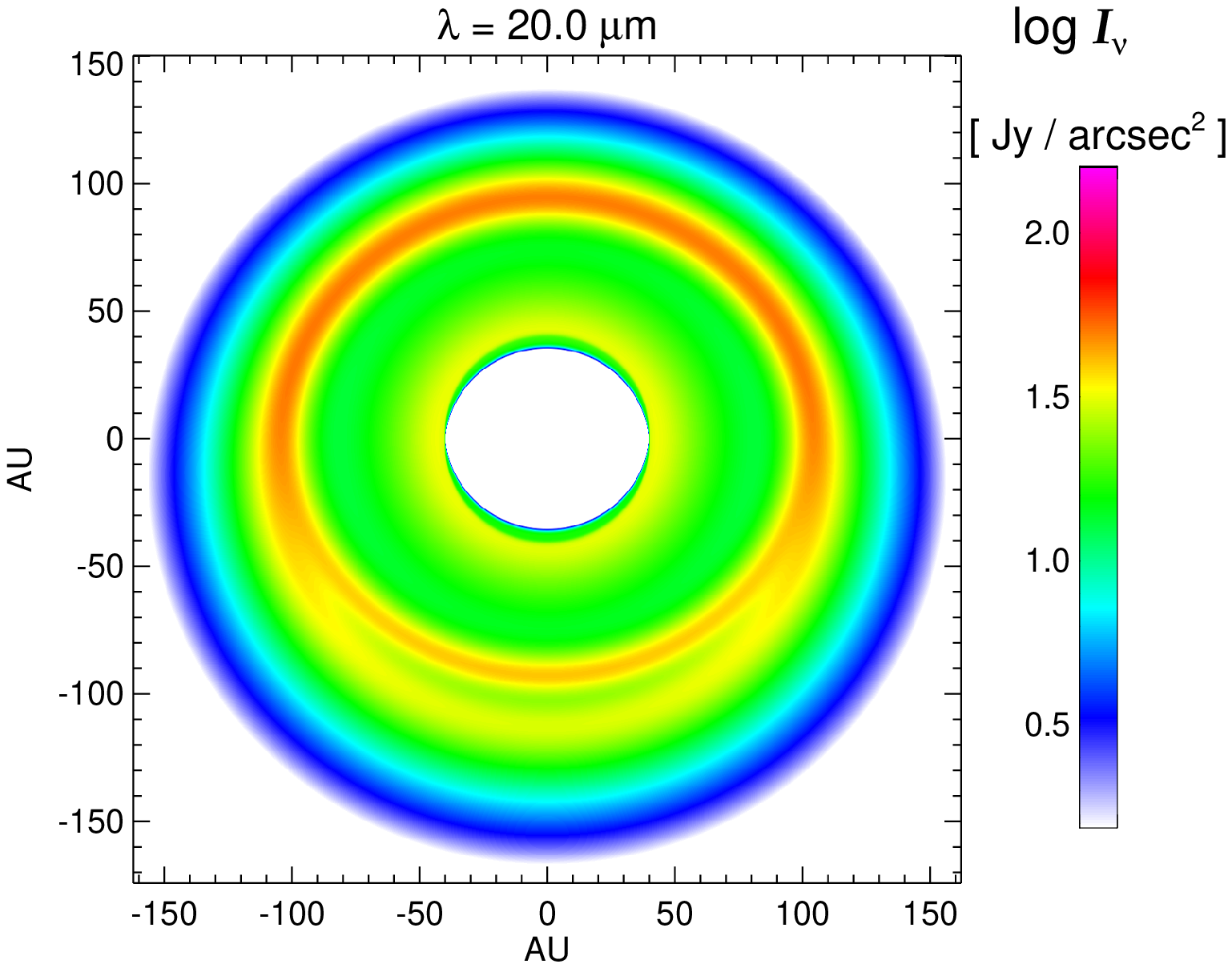}
\end{center}
\caption{Simulated images of the protoplanetary disks with a wall at H and Q bands.
The upper panels are for the model with a sharp edge ($ \Gamma~=~\infty$) while
the lower panels are for the model with a smooth transition ($ \Gamma~=~20$).
The left panels denote the brightness in H band in the logarithmic scale 
while the right panels do those in Q band.} \label{image}
\end{figure}

We obtained simulated images of the protoplanetary disk by integrating
Equation (\ref{transfer0}) along the line of sight.   The source terms are
evaluated from $ T $ and $ E _\nu $ obtained by our M1 model.
The upper panels of Figure~\ref{image} show the simulated images at 
$ \lambda~=~$ 1.58~$\mu$m (left, H-band) and 20~$\mu$m
and (right, Q-band).
The line of sight is assumed to inclined by 15$^\circ$ from the axis normal
to the disk.  Both the images show bright rings at $ r~\simeq~100$~AU.
These images are similar to those of \citet{honda12} who solved the
radiative transfer by the Monte Carlo simulation.

\begin{figure}[h]
\begin{center}
\FigureFile(85mm,85mm){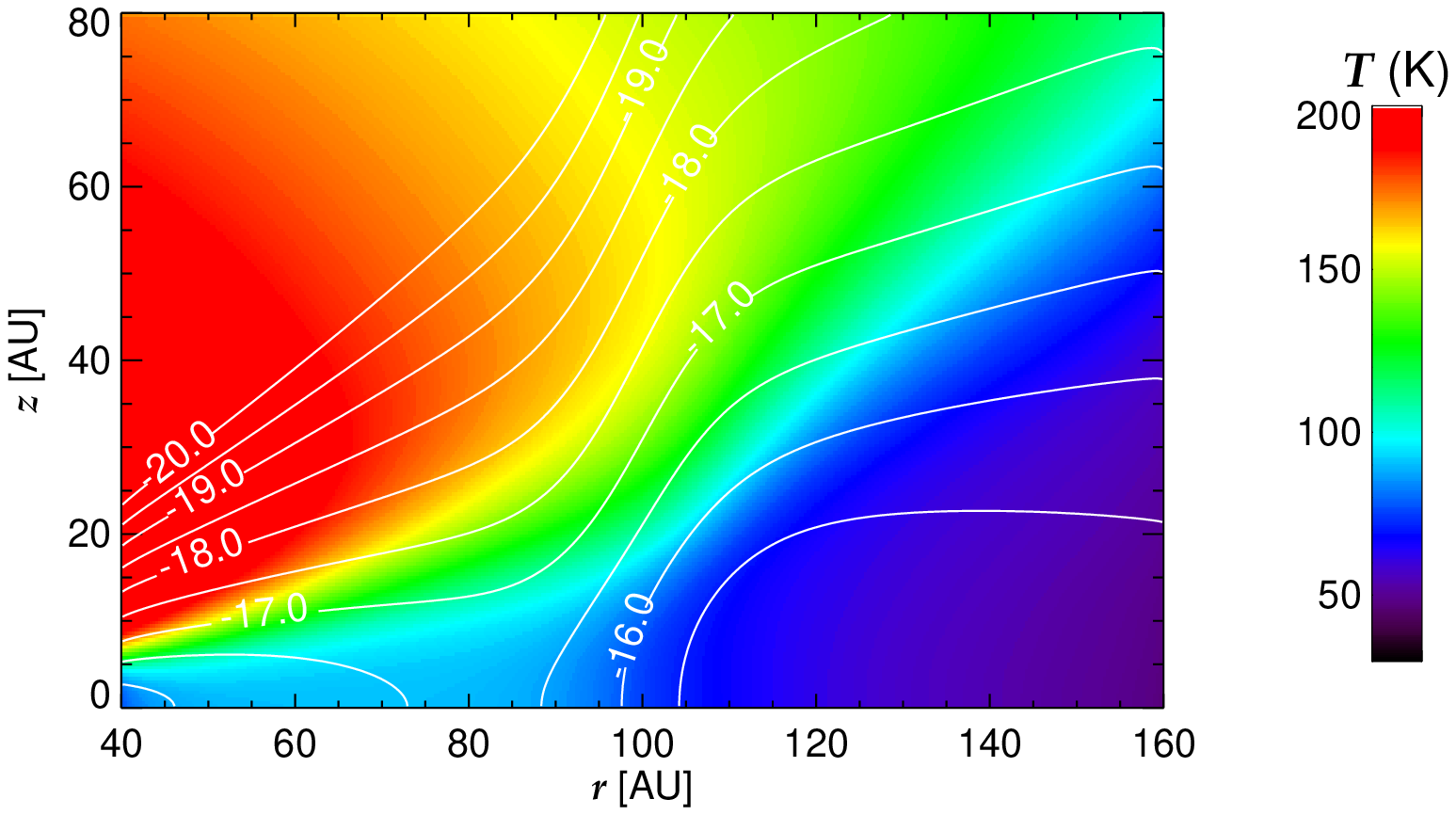}
\end{center}
\caption{The same as Fig.~\ref{2Dsharp_rhoT} but for $\Gamma$~=~20.} 
\label{2Dsmooth_rhoT}
\end{figure}

We made another model by assuming $ \Gamma~=~20 $ while keeping
the other model parameters unchanged.  The temperature and density
distributions are shown in Figure~\ref{2Dsmooth_rhoT}.    The density
change around the wall is smooth and more likely than that shown in
Figure~\ref{2Dsharp_rhoT}.   

\begin{figure}[h]
\begin{center}
\FigureFile(85mm,85mm){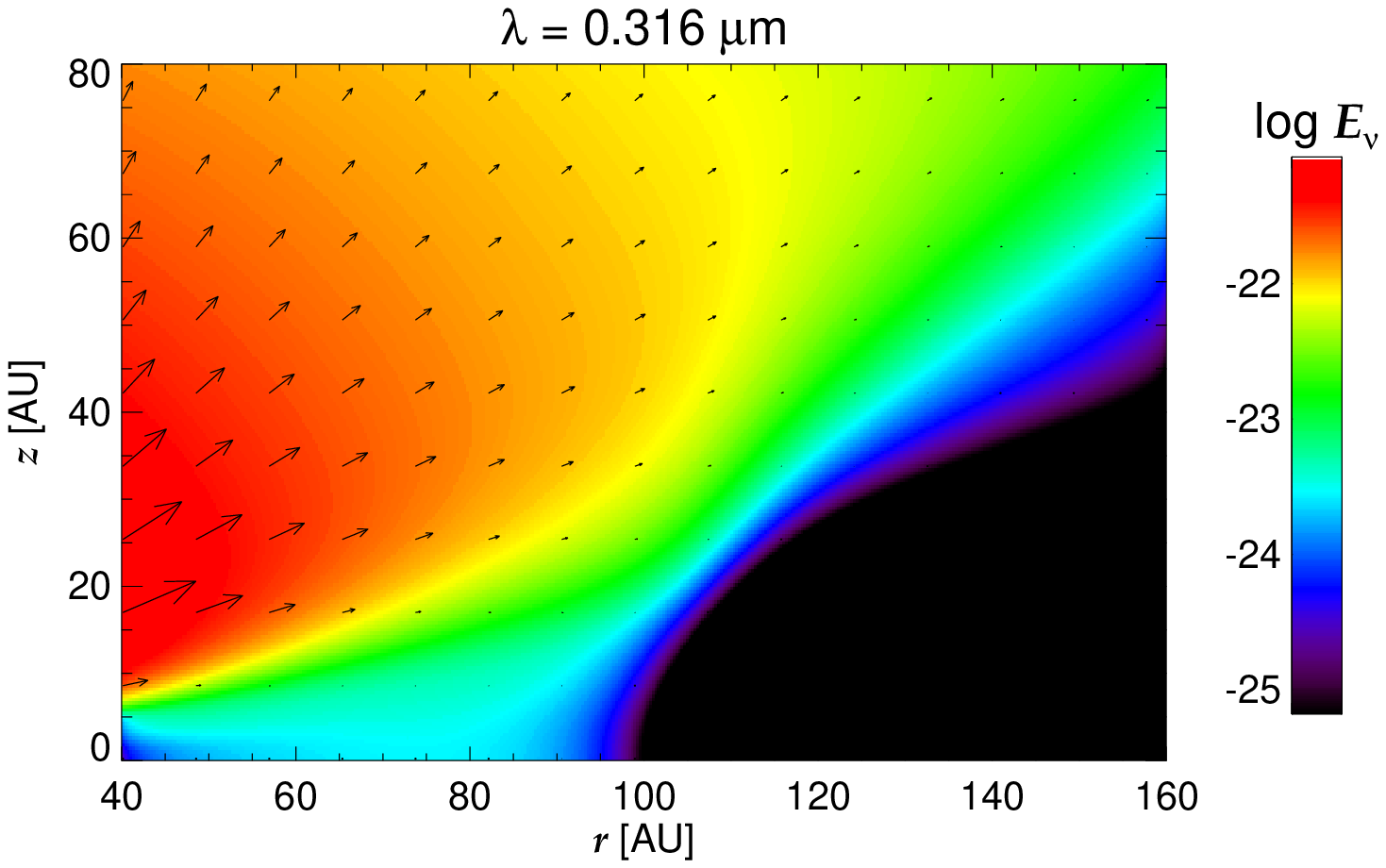} \\
\FigureFile(85mm,85mm){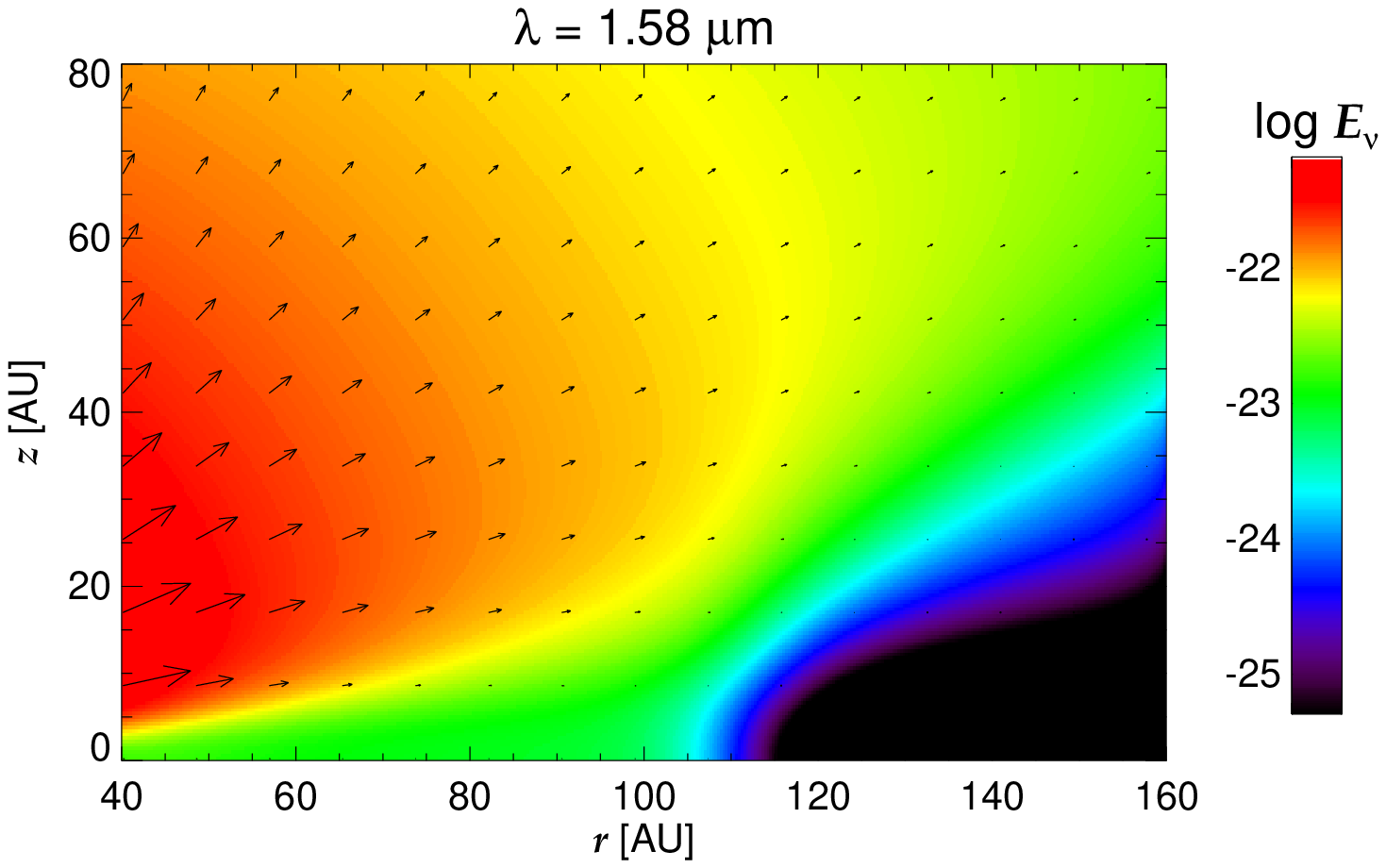} \\
\FigureFile(85mm,85mm){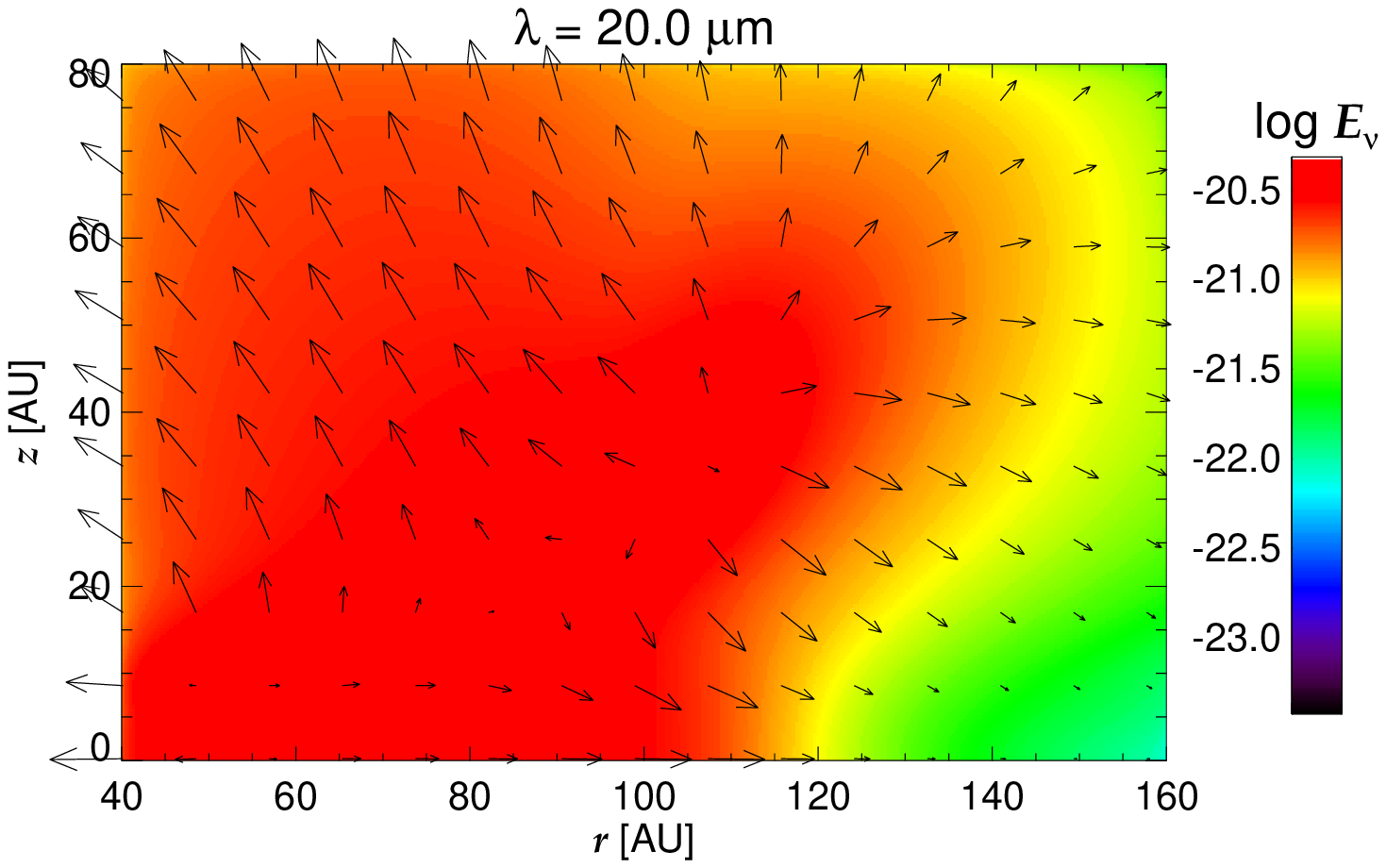} \\
\FigureFile(85mm,85mm){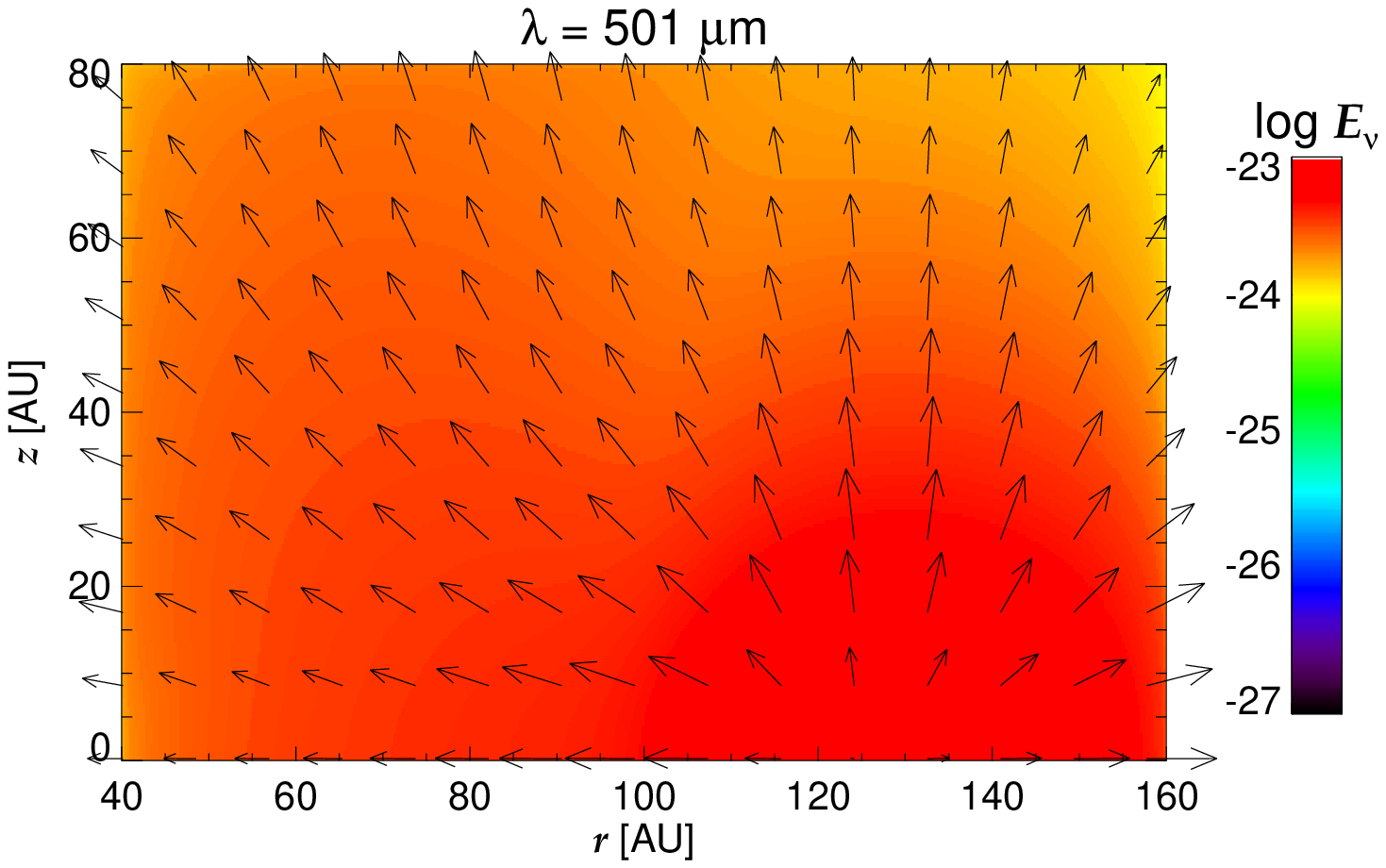}
\end{center}
\caption{The same as Figure \ref{Erad_m0} but for the model having 
$ \Gamma~=~20$.}\label{Erad_m1}
\end{figure}

Figure~\ref{Erad_m1} is the same as Figure~\ref{Erad_m0} but
for $ \Gamma~=~20 $.  The result is almost the same  but the
wall boundary is less sharp as expected.  The photospheres
are located just behind the wall in the model of $ \Gamma~=~\infty$
in a broad range of the wavelength.   However, the location of
the photosphere depends on the wavelength in the model of
$ \Gamma~=~20$.    

The lower panels of Figure~\ref{image} is the same as the upper panels but for 
$ \Gamma~=~20$.   The bright ring are broader in
the model of $ \Gamma~=~20$ than in that of 
$ \Gamma ~=~\infty $.  The peak brightness is lower in
the model of $ \Gamma~=~20 $.  This is because the wall is appreciably
inclined when $ \Gamma~=~20 $ (see Figure~\ref{2Dsmooth_rhoT}).
The ring corresponds the region in which the surface density increases
with the radius.    In other words, the wall is seen as the bright ring in the
image.   Our model will serve to evaluate the surface density
distribution from observed images.

\section{Discussions}

As demonstrated in the previous sections, M1 model works well both in vacuum
and in optically thick media if absorption is taken into account properly in
the numerical flux.   If we use the formal solution for computing time evolution,
the time step can be as large as $ \Delta x/(2c) $ irrespectively of the opitacal
depth.   Thanks to this improvement, we succeeded in applying M1 model to
the protoplanetary disks.   They are optically thick in optical bands while optically 
in far infrared bands.     They are heated by the optical and near infrared stellar
lights and cooled by mid and far infrared emission.   Thus it is important to
take account of both optically thin and thick radiation simultaneously.
Our numerical scheme will expand the applicability of M1 model.   

M1 model can be used also for neutrino transfer.   Neutrino is a major coolant 
in compact objects such as neutron stars and black holes.   It should play an
important role in dynamics of core collapse supernovae and in gamma ray bursts
(see, e.g., a review by \cite{mezzacappa05} and the references therein).
Both core collapse supernovae and gamma ray burst sources are thought to
be highly anisotropic and their dynamics should be studied ideally based
on three dimensional numerical simulations.   It is also essential to
take account of the neutrino energy in the simulations.   Neutrino opacity
is proportional to the square of the neutrino energy.   Low energy neutrinos
are easy to leak while high energy ones not.  Thus M1 model is a  reasonable
choice for numerical simulations of these objects.   It reduces the computation
cost by lowering the angular resolution.   Yet it can express a shadow and 
beamed radiation.  

It should be noted that M1 model can solve the propagation of a flash
by an explicit manner.    Radiation at the next time step depends only
on those in the neighboring cells.   We need neither ray tracing
nor global iteration.   In other words, M1 model does not require global
communication for proceeding a time step.   This is beneficial for 
massive parallel computation, since global communication between
processors are often a bottle neck.

In compact objects, the gas has a high temperature and the sound speed
is comparable to or only by a factor of ten smaller than the speed of light.
Thus it is not serious that the time step is restricted to the light propagation
time, $ 0.5~\Delta x/c $;  a similar small time step is required from the
hydrodynamical simulations if we integrate them explicitly.  If we solve
both radiation and hydrodynamics explicitly, we can take account of 
heating by a flash of neutrino.   The compact sources may have a highly
variable luminosity.  

M1 model may be applied to dynamics of non-relativistic objects in which
the sound speed is much slower than the speed of light.   We can reduce
the speed of light propagation in M1 model to prolong the time step.
If we replace $ c $ by $ c ^\prime $ in Equations (\ref{formalE}) and
(\ref{formalF}), the propagation speed is reduced to $ c ^\prime $.
We expect that the reduction does not affect the result seriously from
the following reason.
Both hydrodynamical and thermal timescales are much longer than
the time for light propagation in most non-relativistic objects.
Thus the light propagation speed is assumed to be infinite in some 
radiation hydrodynamics.   The results are valid since the light
speed is much faster than other speeds of waves  and we can
neglect  the difference between the real light speed and infinity.
We think that we can neglect the difference between $ c $
and $ c ^\prime $ as long as $ c ^\prime $ is much larger than
other speed of wave propagation.  Our idea is based on the
heuristic experiment by \citet{hotta12}.   They performed
numerical simulations of convection  in the Sun by reducing 
the sound speed artificially.   The velocity of material convection
is much lower than the sound speed.  Thus, most of the simulations
thus far apply the anelastic approximation in which the sound
speed is infinitely large.  However, \citet{hotta12} demonstrated
that nearly the same results are obtained even when the sound speed
is reduced artificially as far as the reduced sound speed is still 
much faster than the convection velocity. Their experiment suggests
us that we can reduce the light speed artificially without loss of quality.

In summary, M1 model has potential applicability to many problems 
including protoplanety disks, core collapse supernovae, and gamma ray
bursts.

%%%%%%%%%%%%%%%%%%%%%%%%%%%%%%%%%%%%%%%

\bigskip

We thank Profs.  Mitsuru Honda, Taishi Nakamoto, Hideko Nomura, and Motohide
Tamura  for providing us references on the protoplanetary disks.
We also thank Dr. Hiroyuki Takahashi for showing us his unpublished work
on M1 model and Drs. Edouard Audit, B. Dubroca, and Sebastian 
Fromang for valuable discussions. 
TH hanks Yukawa Institute and Prof. Masaru Shibata for their warm hospitality.
This manuscript was prepared during his stay at Yukawa Insitute, Kyoto University.
He also thank la Maison de la Simulation for their warm hospitality during his stay.
Some preliminary results of this paper were presented at the 7th International
Conference on Numerical Modeling of Space Plasma held in Hawaii in June 2012
and will be included in the proceedings.   
This work is financially supported in part by the Grant-in-Aid for Scientific 
Research on Innovative Areas (23105703).

\appendix

\section*{Numerical Flux of the Second Order Accuracy in Space}

First we introduce a new variable,
\begin{eqnarray}
E _\nu ^\prime & \equiv & \sqrt{\displaystyle
E _\nu ^2 \, - \, \left( \frac{\mbox{\boldmath$F$}_\nu}{c} \right) ^2 } 
\end{eqnarray}
for later convenience.   We evaluate $ E _{\nu,j+1/2} ^\prime $ on the cell boundary
by extrapolation with the minmod limiter and obtain
\begin{eqnarray}
E _{\nu,j+1/2} ^{\rm \prime (L)*} & = &
E ^\prime _{\nu,j} \, + \, \frac{1}{2} \Delta E _{\nu, j+1/2} ^{\rm \prime (L)} , \\
E _{\nu,j+1/2} ^{\rm \prime (R)*} & = &
E ^\prime _{\nu,j} \, - \, \frac{1}{2} \Delta E _{\nu,j+1/2} ^{\rm \prime (R)} , \\
\Delta E _{\nu,j+1/2} ^{\prime \rm (L)} & = & \frac{1}{2} \,
\min \left( \left| \Delta E ^\prime _{\nu,j+1/2}\right|, \,
\left|\Delta E ^\prime _{\nu,j-1/2} \right| \right) \, \nonumber \\
& \; & \times
\left[ {\rm sgn} \left(\Delta E ^\prime _{\nu,j+1/2} \right)
\, + \, {\rm sgn} \left(\Delta E ^\prime _{\nu,j-1/2}
\right) \right]  \, , \\
\Delta E _{\nu,j+1/2} ^{\prime \rm (R)} & = & \frac{1}{2} \,
\min \left( \left| \Delta E ^\prime _{\nu,j+1/2}\right|, \,
\left|\Delta E ^\prime _{\nu,j+3/2} \right| \right) \, \nonumber \\
& \; & \times 
\left[ {\rm sgn} \left(\Delta E ^\prime _{\nu,j+1/2} \right)
\, + \, {\rm sgn} \left(\Delta E ^\prime _{\nu,j+3/2}
\right) \right] , \\
\Delta E ^\prime _{\nu,j-1/2} & = & E ^\prime _{\nu,j} \, - \, E ^\prime_{\nu,j-1} , \\
\Delta E ^\prime _{\nu,j+1/2} & = & E ^\prime _{\nu,j+1} \, - \, E ^\prime _{\nu,j} , \\
\Delta E ^\prime _{\nu,j+3/2} & = & E ^\prime _{\nu,j+2} \, - \, E ^\prime_{\nu,j+1} , 
\end{eqnarray}
where sgn denotes the sign function.  Similarly we evaluate 
$  F _{x,\nu,j+1/2} $ on the cell boundary to obtain
\begin{eqnarray}
F _{x,\nu,j+1/2} ^{\rm (L)*} & = &
F _{x,\nu,j} \, + \, \frac{1}{2} \Delta F _{x,\nu,j+1/2} ^{\rm (L)} \, , \\
F _{x,\nu,j+1/2} ^{\rm (R)*} & = &
F _{x,\nu,j} \, - \, \frac{1}{2} \Delta F _{x,\nu,j+1/2} ^{\rm (R)} \, ,\\
\Delta F _{x,\nu,j+1/2} ^{\rm (L)} & = & \frac{1}{2} \,
\min \left( \left| \Delta F _{x,\nu,j+1/2}\right|, \,
\left|\Delta F _{x,\nu,j-1/2} \right| \right) \nonumber \\
& \;  & \left[ {\rm sgn} \left(\Delta F _{x,\nu,j+1/2} \right)
\, + \, {\rm sgn} \left(\Delta F _{x,\nu,j-1/2} \right) \right] \, , \\
\Delta F _{x,\nu,j+1/2} ^{\rm (R)} & = & \frac{1}{2} 
\min \left( \left| \Delta F _{x,\nu,j+1/2}\right|, \,
\left|\Delta F _{x,\nu,j+3/2} \right| \right) \nonumber \\
& \; &
\left[ {\rm sgn} \left(\Delta F _{x,j+1/2} \right)
\, + \, {\rm sgn} \left(\Delta F _{x,j+3/2}
\right) \right] \, , \\
\Delta F _{x,\nu,j-1/2} & = & F _{x,\nu,j} \, - \, F _{x,\nu,j-1} , \\
\Delta F _{x,\nu,j+1/2} & = & F _{x,\nu,j+1} \, - \, F _{x,\nu,j} ,\\
\Delta F _{x,\nu,j+3/2} & = & F _{x,\nu,j+2} \, - \, F _{x,\nu,j+1} .
\end{eqnarray}
We obtain $ F _{y,\nu,j+1/2} ^{\rm (L)*} $,  $ F _{y,\nu,j+1/2} ^{\rm (R)*} $, 
$ F _{z,\nu,j+1/2} ^{\rm (L)*} $ and $ F _{z,\nu,j+1/2} ^{\rm (R)*} $ by the
same procedure.   The energy density on the cell boundary is evaluated
to be
\begin{eqnarray}
E _{\nu,j+1/2} ^{\rm (L)*} & = &
\sqrt{ \displaystyle \left( E _{\nu,j+1/2} ^{\prime \rm (L)*} \right) ^2
\, + \, \frac{\displaystyle  
\left[ \left( F _{x,\nu,j+1/2} ^{\rm (L)*} \right) ^2 +
\left( F _{y,\nu,j+1/2} ^{\rm (L)*} \right) ^2 +
\left( F _{z,\nu,j+1/2} ^{\rm (L)*} \right) ^2
\right]}{c ^2} } ,  \\
E _{\nu,j+1/2} ^{\rm (R)*} & = &
\sqrt{ \displaystyle \left( E _{\nu,j+1/2} ^{\prime \rm (R)*} \right) ^2
\, + \, \frac{\displaystyle  
\left[ \left( F _{x,\nu,j+1/2} ^{\rm (R)*} \right) ^2 +
\left( F _{y,\nu,j+1/2} ^{\rm (R)*} \right) ^2 +
\left( F _{z,\nu,j+1/2} ^{\rm (R)*} \right) ^2
\right]}{c ^2} } .
\end{eqnarray}
Introduction of  $ E _\nu ^\prime $ guarantees that the
energy density is larger than the energy flux divided the
speed of light on the cell boundary.

The energy density, $ E _{\nu,j+1/2} ^{\rm (L)*} $, can exceed
$ (3/2) E _{\nu,j}  $.   If we would apply MUSCL to $ E _\nu $, 
it can not so high.   To avoid such a high energy
density,  we reduce both the energy
density and flux by multiplying the factors,
\begin{eqnarray}
\varepsilon _\nu ^{\rm (L)} & = & \left\{ 
\begin{array}{ll}
1 & {\rm if} \; E _{\nu,j+1/2} ^{\rm (L)*} \, \le \, \displaystyle \frac{3}{2} E _{\nu,j} \\
\displaystyle \frac{3}{2} \, \frac{E _{\nu,j}}{E _{\nu,j+1/2} ^{\rm (L)*}} &
{\rm otherwise}
\end{array}  \right. \, , \\
\varepsilon _\nu ^{\rm (R)} & = & \left\{
\begin{array}{ll}
1 & {\rm if} \; E _{\nu,j+1/2} ^{\rm (R)*} \, \le \, \displaystyle \frac{3}{2} E _{\nu,j+1} \\
\displaystyle \frac{3}{2} \, \frac{E _{\nu,j+1}}{E _{\nu,j+1/2} ^{\rm (R)*}} &
{\rm otherwise}
\end{array}  \right. .
\end{eqnarray}
Accordingly we obtain
\begin{eqnarray}
E _{\nu,j+1/2} ^{\rm (L)} & = & \varepsilon  _\nu ^{\rm (L)} \,
E _{\nu,j+1/2} ^{\rm (L)*} , \\
\mbox{\boldmath$F$} _{\nu,j+1/2} ^{\rm (L)} & = & \varepsilon _\nu ^{\rm (L)} \,
\mbox{\boldmath$F$} _{\nu,j+1/2} ^{\rm (L)*} , \\
E _{\nu,j+1/2} ^{\rm (R)} & = & \varepsilon  _\nu ^{\rm (R)} \,
E _{\nu,j+1/2} ^{\rm (R)*} , \\
\mbox{\boldmath$F$} _{\nu,j+1/2} ^{\rm (R)} & = & \varepsilon _\nu ^{\rm (R)} \,
\mbox{\boldmath$F$} _{\nu,j+1/2} ^{\rm (R)*} . 
\end{eqnarray}
We use $ E _{\nu,j+1/2} ^{\rm (L)} $ and $ \mbox{\boldmath$F$} _{\nu,j+1/2} 
^{\rm (L)} $ for computing the energy flux from cell $ j $ to $ j +1 $, and
$ E _{\nu,j+1/2} ^{\rm (R)} $ and $ \mbox{\boldmath$F$} _{\nu,j+1/2} 
^{\rm (R)} $ for that from $ j+1$ to $ j $.

Emission and scattering within a numerical cell are included in the
second order numerical flux by the following procedure.
First we evaluate the the blackbody emission on the cell boundary,
$ B _{\nu,j+1/2} ^{\rm (L)} $ and $  B _{\nu,j+1/2} ^{\rm (R)} $ by the MUSCL approach,
\begin{eqnarray}
B _{\nu,j+1/2} ^{\rm (L)} & = & B _{\nu,j} \, + \, \frac{1}{2}
\, \Delta B _{\nu,j+1/2} ^{\rm (L)} \, , \\
B _{\nu,j+1/2} ^{\rm (R)} & = & B _{\nu,j+1} \, - \, \frac{1}{2}
\, \Delta B _{\nu,j+1/2} ^{\rm (R)} \, ,\\
\Delta B _{\nu,j+1/2} ^{\rm (L)} & = &
\min \, \left( | \Delta B _{\nu,j+1/2} |,~|\Delta B _{\nu,j-1/2}|
\right) \nonumber \\
& \; & \times \left[ {\rm sgn} \left( \frac{1}{2}, \, \Delta B _{\nu,j+1/2}
\right) \, + \, {\rm sgn} \left( \frac{1}{2}, \,
\Delta B _{\nu,j-1/2} \right) \right]  \, ,\\
\Delta B _{\nu,j+1/2} ^{\rm (R)} & = &
\min \, \left( | \Delta B _{\nu,j+1/2} |,~|\Delta B _{\nu,j+3/2}|
\right) \nonumber \\
& \; & \times \left[ {\rm sgn} \left( \frac{1}{2}, \, \Delta B _{\nu,j+1/2}
\right) \, + \, {\rm sgn} \left( \frac{1}{2}, \,
\Delta B _{\nu,j+3/2} \right) \right] \, .
\end{eqnarray}
We assume that $ B _\nu $ is a linear function of the optical
depth between the cell center and boundary.   Then we obtain
\begin{eqnarray}
\int _{x_{j}} ^{x_{j+1/2}} \kappa _{\nu,a,j} \rho _j B _\nu (x^\prime) \, dx ^\prime &
= & B _{\nu,j+1/2} ^{\rm (L)} \,  \left[ 1 \, - \, \frac{1 \, - \, e ^{-w _{\nu,j}}}{w _{\nu,j}} \right]
\nonumber \\
&  + & B _{\nu,j} \, \left[   \frac{1 \, -  \, e ^{-w _{\nu,j}}}{w _{\nu,j}} 
\, - \, e ^{-w _{\nu,j}} \right] , 
\end{eqnarray}
where
\begin{equation}
w _{\nu,j} \; = \;  \kappa _{\nu,a} \rho \, \Delta x _j \, .
\end{equation}
The symbol $ \Delta x _j $ denotes the cell width.
Similarly we can evaluate the scattering within the cell.

By taking the emission
and scattering within the cell, the numerical flux of the second
order accuracy is expressed as 
\begin{eqnarray}
F _{\nu,j+1/2} ^* & = & \zeta _{\nu,j} 
F _{\nu,j+1/2} ^{\rm (L)} 
\; + \; \eta _{\nu,j} ^\prime
\pi B _{\nu,j+1/2} ^{\rm (L)} \, + \,
\eta _{\nu,j} ^{\prime\prime}
\pi B _{\nu,j} \nonumber \\
& \; & + 
\eta _{\nu,j} \,
\left[
\zeta _{\nu,j} ^\prime \, 
\frac{c E _{\nu,j+1/2} ^{\rm (L)}}{4} \, + \, \zeta _{\nu,j} 
^{\prime\prime} \frac{c E _{\nu,j}}{4} \right] \nonumber \\
& \; & + \, \zeta _{\nu,j+1} 
F _{\nu,j+1/2} ^{\rm (R)} 
\; - \; \eta _{\nu,j+1} ^\prime
\pi B _{\nu,j+1/2} ^{\rm (R)} \, - \,
\eta _{\nu,j+1} ^{\prime\prime}
\pi B _{\nu,j+1} \nonumber \\
& \; & - \eta _{\nu,j+1} \,
\left[
\zeta _{\nu,j+1} ^\prime \, 
\frac{c E _{\nu,j+1/2} ^{\rm (R)}}{4} \, + \, \zeta _{\nu,j+1} 
^{\prime\prime} \frac{c E _{\nu,j+1}}{4} \right] ,
\end{eqnarray}
\begin{eqnarray}
P _{\nu,ii,j+1/2} ^* & = & \zeta _{\nu,j} 
P _{\nu,ii,j+1/2} ^{\rm (L)} 
\; + \; \eta _{\nu,j} ^\prime
\frac{2\pi}{3c} B _{\nu,j+1/2} ^{\rm (L)} \, + \,
\eta _{\nu,j} ^{\prime\prime}
\frac{2\pi}{3c} B _{\nu,j} \nonumber \\
& \; & + \,
\eta _{\nu,j} \,
\left[
\zeta _{\nu,j} ^\prime \, 
\frac{E _{\nu,j+1/2} ^{\rm (L)} }{6} \, + \, \zeta _{\nu,j} 
^{\prime\prime} \frac{E _{\nu,j}}{6} \right] \nonumber \\
& \; & + \, \zeta _{\nu,j+1} 
P _{\nu,ii,j+1/2} ^{\rm (R)} 
\; + \; \eta _{\nu,j+1} ^\prime
\frac{2\pi}{3c} B _{\nu,j+1/2} ^{\rm (R)} \, + \,
\eta _{\nu,j+1} ^{\prime\prime}
\frac{2\pi}{3c} B _{\nu,j+1} \nonumber \\
& \; & + \, \eta _{\nu,j+1} \,
\left[
\zeta _{\nu,j+1} ^\prime \, 
\frac{E _{\nu,j+1/2} ^{\rm (R)}}{6} \, + \, \zeta _{\nu,j+1} 
^{\prime\prime} \frac{c E _{\nu,j+1}}{6} \right] , \\
P _{\nu,ik,j+1/2} ^* & = &
\eta _{\nu,j} P _{\nu,ik,j+1/2} ^{\rm (L)} \; + \;
\eta _{\nu,j+1} P _{\nu,ik,j+1/2} ^{\rm (R)} \; 
\hskip 1.0cm
{\rm if} \; i~\ne~k ,
\end{eqnarray}
\begin{eqnarray}
\eta _{\nu,j} & = & e ^{-w_{\nu,j}} , \\
\eta _{\nu,j} ^\prime & = &
\left[ 1 \, - \, \frac{1 \, - \, e ^{-w_{\nu,j}}}
{w _{\nu,j}} \right] ,\\
\eta _{\nu,j} ^{\prime\prime} & = &
\left[\frac{1 \, - \, e ^{-w_{\nu,j}}}
{w _{\nu,j}} \, - \, e ^{-w _{\nu,j}} \right]  ,\\
% w _{\nu,j} & = & \frac{1}{2} \kappa _{\nu,a,j} \rho _j
% \Delta x_j ,\\
\zeta _{\nu,j} & = & e ^{-w_{\nu,j} \, - \, s _{\nu,j}} ,\\
\zeta _{\nu,j} ^\prime & = &
\left[ 1 \, - \, \frac{1 \, - \, e ^{-s_{\nu,j}}} {s  _{\nu,j}} \right]  ,\\
\zeta _{\nu,j} ^{\prime\prime} & = &
\left[\frac{1 \, - \, e ^{-s_{\nu,j}} }
{s _{\nu,j}} \, - \, e ^{-s _{\nu,j} } \right]  , \\
s _{\nu,j}  & = & \frac{1}{2} \kappa _{\nu,s,j} \rho _j 
\Delta x _j .
\end{eqnarray}

%%%
% See the manual for the detail.
%%%


\begin{thebibliography}{}
% Journals(e.g. A\&A,ApJ,AJ,NMRAS,PASP ...)
% Authors, Year, Journal, Vol#, Page#
% Journal Title Abbreviation >> http://www.asj.or.jp/pasj/Jabb.html
\bibitem[Audit et al.(2002)]{audit02} Audit, E., Charrier, P., Chi\`eze, J.-P., 
\& Dubroca, B. 2002, astro-ph/0206281v1
\bibitem[Berthon, Charrier, \& Dubroca(2007)]{berthon07}
Berthon, C., Charrier, P., \& Dubroca, B. 2007, J. Sci. Comput., 31, 347
\bibitem[Castor(2004)]{castor04}
Castor, J. 2004, Radiation Hydrodynamics, (Cambridge Univ., Cambridge) 
\bibitem[Chiang \& Goldreich(1997)]{chiang97} Chiang, E.~I., \& Goldreich, P.
1997, \apj, 490, 368
\bibitem[D'Alessio et al.(1998)]{dalessio98}  D'Alessio, P., Cant\'o, J., Calvet, N., \&
Lizano, S. 1998, \apj, 500, 411
\bibitem[Deschpande(1986)]{deschpande86} Deshpande, S.M. 1986,
in the proceedings of  AIAA 24th Aerospace Science Meeting (American Institute 
of Aeronautics and Astronautics, New York) Paper 86-0275 
\bibitem[Draine(2003)]{draine03} Draine, B.~T. 2003, \araa, 41, 241
\bibitem[Dubroca et al.(2003)] {dubroca03} Dubroca, B., 
Frank, M., Klar, A., \& Th\"ommes, G. 2003, Z.  Angew. Math. Mech., 83, 853
\bibitem[Dubroca \& Feugueas(1999)]{dubroca99} 
Dubroca, B., \& Feugeas, J.~L. 1999, CRAS, 329, 915
\bibitem[Dullemond, Dominki, \& Natta(2001)]{dullemond01} 
Dullemond, C.~P., Dominik, C., \& Natta, A. 2001, \apj, 560, 957
\bibitem[Fukagawa et al.(2004)]{fukagawa04} Fukagawa, M. et al.  2004, \apj, 
605, 53
\bibitem[Gonz\'alez et al.(2007)]{gonzalez07}
Gonz\'alez,  M., Audit, E., \& Huyunh,  P. 2007, \aap, 464, 429 
\bibitem[Grady et al.(1997)]{grady99} Grady, C. A., Woodgate, B., Bruhweiler, F. C., 
Boggess, A., Plait, P., Lindler, D. J., Clampin, M., \& Kalas, P. 1999, \apjlett, 523, L151
\bibitem[Hashimoto et al.(2011)]{hashimoto11} Hashimoto, J. et al. 2011,
\apjlett, 729, 17
\bibitem[Henning \& Stognienko(1996)]{henning96} Henning, T. \& Stognienko, R.
1996, \aap, 311, 291
\bibitem[Hauck(2011)]{hauck11} Hauck. C.~D. 2011,
Commun. Math. Sci. 9, 187
\bibitem[Hirsch(1990)]{hirsch90} Hirsch, C. 1990, Numerical Computation of 
Internal and External Flows, Vol. 2 (Chichester: Wiley) 
\bibitem[Honda et al.(2012)]{honda12} Honda, M. et al. 2012, \apj, 752, 143 
\bibitem[Hotta et al.(2012)]{hotta12} Hotta, H., Rempel, M. Yokoyama, T.,
Iida, Y., \& Fan, Y. 2012, \aap, 539, 30
\bibitem[Levermore(1984)]{levermore84}
Levermore, C.~D. 1984, JQSRT, 31, 149
\bibitem[Mezzacappa(2005)]{mezzacappa05} Mezzacappa, A. 2005,
Ann. Rev. Nucl. Part. Sci., 55, 467
\bibitem[Okamoto, Yoshikawa  \& Umemura(2012)]{okamoto12} Okamoto, T.,
Yoshikawa,  K.,  \& Umemura, M. 2012, \mnras, 419, 2855
\bibitem[Pullin(1980)]{pullin80} Pullin, D.~I. J. Comp. Phys., 34, 231
\bibitem[Roe(1981)]{roe81} Roe, P.~L. 1981, J. Comp. Phys., 43. 357
\bibitem[Susa(2006)]{susa06} Susa, H. 2006, \pasj, 58, 445 
\bibitem[Toro(2009)]{toro09}  Toro, E.~F. 2009, Riemann Solvers
and Numerical Methods for Fluid Dynamics, 3rd Ed. (Springer,
Berlin) Ch. 6
\bibitem[van den Ancker et al.(1997)]{vandenacker97}
van den Ancker, M. E., The, P. S., Tjin A Djie, H. R. E., Catala, C., 
de Winter, D., Blondel, P. F. C., \& Waters, L. B. F. M. 1997, \aap, 324, L33
\bibitem[Williams \& Cieza(2011)]{williams11} Williams, J.~P., \& Cieza, L.~A.
2011, \araa, 49, 67
\end{thebibliography}
\end{document}